\documentclass[aip,
amsmath,
amssymb,
preprint,
reprint,
]{revtex4-1}

\usepackage{graphicx}
\usepackage{dcolumn}
\usepackage{bm}

\usepackage[utf8]{inputenc}
\usepackage[T1]{fontenc}
\usepackage{mathptmx}
\usepackage{etoolbox}

\makeatletter
\def\@email#1#2{%
 \endgroup
 \patchcmd{\titleblock@produce}
  {\frontmatter@RRAPformat}
  {\frontmatter@RRAPformat{\produce@RRAP{*#1\href{mailto:#2}{#2}}}\frontmatter@RRAPformat}
  {}{}
}%
\makeatother
\begin{document}

\preprint{AIP/123-QED}

\title[Coupling phase interference effects in a multimode cavity magnonics system]{Coupling phase interference effects in \\a multimode cavity magnonics system}
\author{M. Avicena}
 \affiliation{ 
IMT Atlantique, Lab-STICC, UMR CNRS 6285, F-29238 Brest, France.
}%
 \email{mufti.avicena@imt-atlantique.fr}
\author{G. Bourcin}%
\affiliation{ 
kwan-tek, 1 rue Galilée, Espace Innova, 56270 Ploemeur, France
}%

\author{V. Vlaminck}
\affiliation{ 
IMT Atlantique, Lab-STICC, UMR CNRS 6285, F-29238 Brest, France.
}%
\author{J. Bourhill}
 \affiliation{ 
Quantum Technologies and Dark Matter Labs, Department of Physics, University of Western Australia, 35 Stirling Hwy, 6009 Crawley, Western Australia.
}%
\author{V. Castel}
\affiliation{ 
IMT Atlantique, Lab-STICC, UMR CNRS 6285, F-29238 Brest, France.
}%
 \email{vincent.castel@imt-atlantique.fr}

\newcommand{\dv}{\mathrm{d}}
\newcommand{\bck}{$S_{12}$}
\newcommand{\fwd}{$S_{21}$}

\date{\today}

\begin{abstract}
Coupling phases play a decisive yet often overlooked role in cavity magnonics, particularly in complex multimode systems. Here, we investigate phase-mediated interference effects in a cavity magnonics system comprising a four-post re-entrant microwave cavity coupled to Yttrium Iron Garnet (YIG) spheres. Using an input–output model that explicitly accounts for both internal and external coupling phases, we achieve agreement with experimental microwave transmission measurements. Our results unravel the emergence of a positionally-dependent uncoupled mode due to interference of cavity photon-magnon (internal) coupling phases. Further, we experimentally observed large nonreciprocity at the antiresonance frequencies and show that this feature arises due to the far-detuned modes' odd parity cavity photon-probe (external) coupling phase interfering with the internal coupling phases. Bridging theory, simulation and experiment, these results establish coupling-phase engineering as a key principle for accurately modeling and designing multimode cavity magnonics devices. 
\end{abstract}

\maketitle

\begin{quotation}
In this work, we explain the origin of a position-selective coupling to degenerate discrete whispering-gallery-mode-like modes, and nonreciprocal transmission for special YIG arrangements through the lens of coupling phases. We show in this work that the selective coupling can be traced to interference effects of internal coupling phases, while the nonreciprocal transmission to be caused by the interference between internal and external coupling phases of far detuned modes. The results presented in this work establish coupling-phase engineering as a key principle for accurately modelling and designing cavity magnonics devices.
\end{quotation}

\section{\label{sec:introduction} Introduction}
Cavity magnonics is an emerging field investigating the interaction between magnons and cavity photons. This interaction can reach the strong-coupling regime and beyond, forming hybridized bosonic quasiparticles known as cavity-magnon polaritons (CMPs). Since its first theoretical prediction \cite{soykalStrongFieldInteractions2010} and early experimental realizations at both cryogenic \cite{tabuchiHybridizingFerromagneticMagnons2014, hueblHighCooperativityCoupled2013, zhangStronglyCoupledMagnons2014} and room temperatures \cite{baiControlMagnonPhoton2016, zhangStronglyCoupledMagnons2014, goryachevHighCooperativityCavityQED2014}, cavity magnonics has been extensively studied, covering topics such as coherent coupling beyond the strong coupling regime \cite{bourhillUltrahighCooperativityInteractions2016, bourcinStrongUltrastrongCoherent2023, golovchanskiyApproachingDeepStrongOnChip2021}, dissipative coupling and level attraction \cite{gardinLevelAttractionInterference2025, harderLevelAttractionDue2018, bourcinLevelAttractionQuasiclosed2024, haoSelectivelyDissipativeCoherent2025}, and nonreciprocity \cite{zhaoPhaseControlledPathwayInterferences2021, raoInterferometricControlMagnoninduced2021, zhangBroadbandNonreciprocityEnabled2020, liaoTunableinducedTransparencyFast2025}. It holds broad technological potential ranging from quantum sensors \cite{lachance-quirionHybridQuantumSystems2019}, quantum memories \cite{zhangMagnonDarkModes2015}, and dark matter detection \cite{lachance-quirionHybridQuantumSystems2019, zarerameshtiCavityMagnonics2022} among others.

One important direction in cavity magnonics is realizing nonreciprocal devices as it is a key requirement for quantum computing architectures and microwave technologies, enabling essential functions such as signal routing, isolation and source protection \cite{kimNonreciprocityCavityMagnonics2024, zhangBroadbandNonreciprocityEnabled2020, wangNonreciprocityUnidirectionalInvisibility2019}. Currently, ferrimagnetic materials are used ubiquitously as field-tunable non-reciprocal components in nonreciprocal microwave devices. This is mainly due to the antisymmetric property of the Polder susceptibility tensor which selectively couples only to right-handed circularly polarized ac magnetic fields relative to the bias field \cite{ranzaniCirculatorsQuantumLimit2019, polderVIIITheoryFerromagnetic1949, gurevichMagnetizationOscillationsWaves2020}. In cavity magnonics, nonreciprocity has been demonstrated through various mechanisms, including tuning dissipative and coherent magnon-photon coupling \cite{wangNonreciprocityUnidirectionalInvisibility2019}, phase-controlled \cite{bourcinLevelAttractionQuasiclosed2024}, and the use of chiral modes in circular \cite{yuCirculatingCavityMagnon2020, bourhillGenerationCirculatingCavity2023} or square cavities \cite{zhangBroadbandNonreciprocityEnabled2020}. Despite these advances, most theoretical and experimental studies rely on reduced-mode descriptions, where only a few cavity photon modes are retained neglecting the far-detuned modes. While such approaches successfully capture level repulsion and mode hybridization, interference-driven phenomena such as transmission antiresonances \cite{smithExceptionalAntimodesMultiDrive, gardinLevelAttractionInterference2025, boventerSteeringLevelRepulsion2019, harderSpinDynamicalPhase2016} (a narrow dip in signal transmission) and phase-sensitive nonreciprocity still need to be treated. This fact is more pronounced in multimode cavity magnonics architectures.

Previously, Gardin et al. \cite{gardinEngineeringSyntheticGauge2024, gardinManifestationCouplingPhase2023} demonstrated that internal coupling phase (attributed to the direction of the microwave (MW) magnetic field vector inside a ferromagnetic sphere) needs to be considered to accurately determine whether magnon signals in the dispersive regime will cross. In another work, Bourcin et al. \cite{bourcinLevelAttractionQuasiclosed2024} showed the importance of attractive and repulsive contribution of far-detuned cavity modes to explain antiresonance level attraction in a quasi-closed cavity, representing it as an external coupling phase factor which was discussed in prior work by Bourhill et al. \cite{bourhillGenerationCirculatingCavity2023} and Zhang et al. \cite{zhangBroadbandNonreciprocityEnabled2020}. A later work by Wang and Xiao \cite{wangInterpretingSparameterSpectra2025} also discussed the role of probing configuration to the observation of attractive/repulsive coupling in a coupled micro-ring resonators system. However, the origin of nonreciprocal transmission at the antiresonance frequencies still needs further investigation \cite{j.w.raoLevelAttractionLevel2019, bourcinLevelAttractionQuasiclosed2024, bourcinMultimodeInputoutputModel2026}. Building upon these prior works, we combine experimental measurements and finite element (FE) simulations to demonstrate the decisive role of coupling phases in shaping the transmission spectra. We show that cavity modes located far from the CMPs, cannot be neglected. This is because the far-detuned modes' phase contributions induce local interference effects. We have derived in this work a theoretical model based on the input-output formalism \cite{gardinerInputOutputDamped1985} of a cavity magnonics system comprised of a quadruple-post re-entrant cavity and multiple YIG spheres placed at symmetry-breaking positions.

This paper is organized as follows, starting in Sec. \ref{sec:cavdes}, we describe the multimode cavity magnonics system. This includes description of the multiple-post re-entrant cavity, and important parameters such as the coupling strength, and the relevant cavity modes that are incorporated into the model. Next, in Sec. \ref{sec:theory} we introduce the notion of internal and external coupling phase and proceed to give an expression for the scattering parameter (S-parameter) based on the input-output formalism. Sec. \ref{sec:results} presents two main results, the distinct transmission signatures and nonreciprocity at the antiresonance frequencies. For both results, we elucidate their physical origin in terms of coupling phase interferences. Finally, Sec. \ref{sec:conc} concludes the paper and discusses implications for phase-engineered nonreciprocal magnonics devices.

\section{\label{sec:cavdes} Cavity Magnonics System}

\subsection{\label{subsec:fourpost}Quadruple-post re-entrant cavity}

\begin{figure}[h]
\centering
    \includegraphics[width=0.95\linewidth]{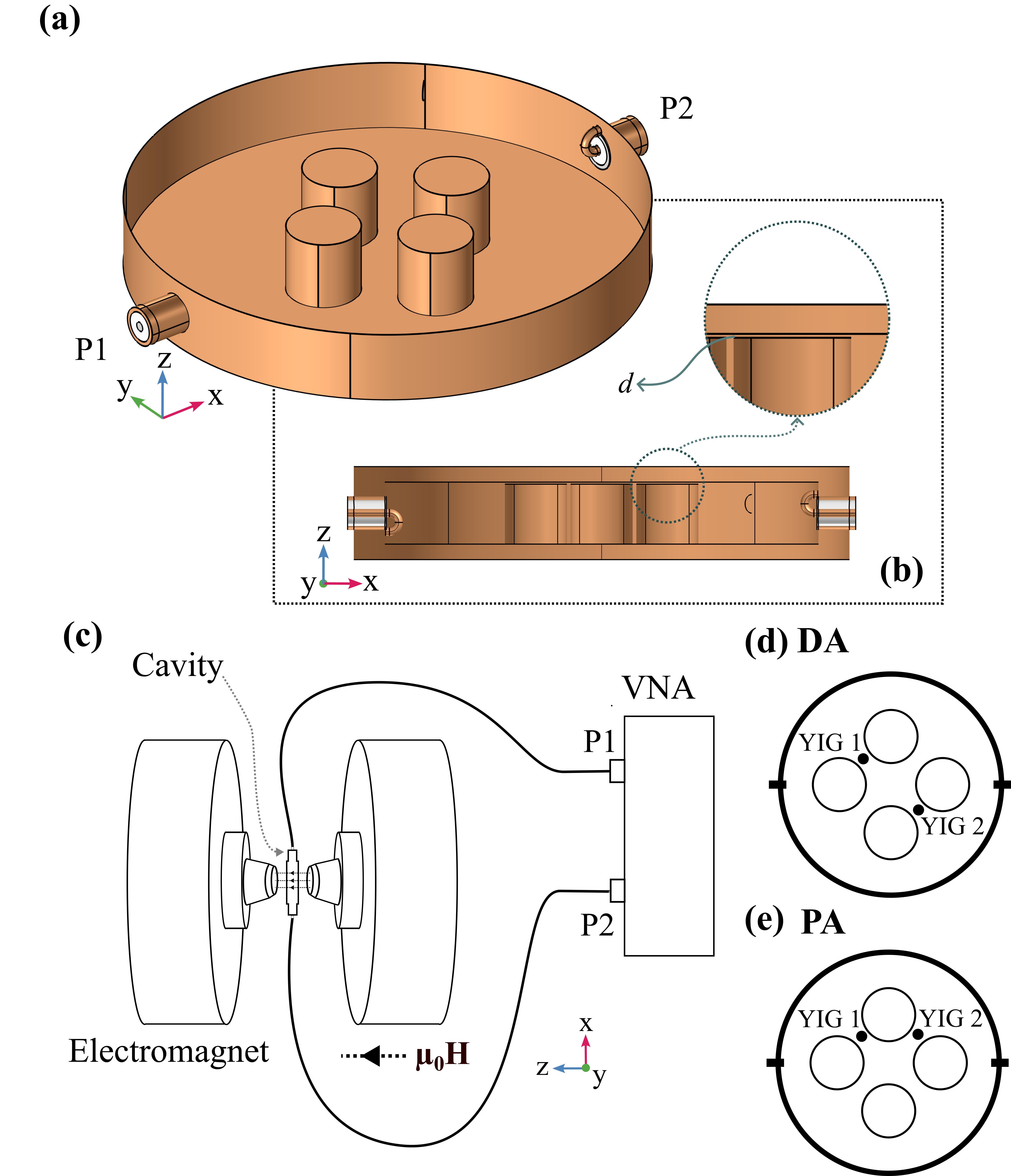}
    \caption{(a) An overhead view of the quadruple-post re-entrant cavity with ports at opposite ends of the cavity. We indicate one end as P1 and P2 as the other. (b) A cross section of the cavity along the zx-plane. The resonance frequency is controlled by the geometry of the cavity, mainly by the lid-post gap distance, $d$ shown in the zoomed in circle. (c) The experimental setup, where the cavity magnonic system is placed so that the posts' orientation is parallel to the external magnetic field, $\mu_0 H$; i.e. we set the z-axis to coincide with $\mu_0 H$. The cavity modes are excited and probed using a vector network analyzer (VNA). (d) Overhead sketch of the diagonal arrangement, DA and (e) sketch of the parallel arrangement, PA.}
    \label{fig:cavity}
\end{figure}

The quadruple-post re-entrant cavity resonator used in this work was modeled using COMSOL Multiphysics\textsuperscript{\tiny\textregistered} and manufactured entirely in-house, employing CNC machining on brass. This type of cavity is an extension of multiple-post re-entrant cavities that have been employed in previous studies \cite{fujisawaGeneralTreatmentKlystron1958, leflochRigorousAnalysisHighly2013, eshbachSpinWavePropagationMagnetoelastic1962, goryachevHighCooperativityCavityQED2014, bourhillUniversalCharacterizationCavity2020}. As shown in Figure \ref{fig:cavity}(a, b), the cavity features four cylindrical posts arranged symmetrically around the center of the circular structure. Two magnetic loop probes, formed from 1.19 mm coaxial cable cores are inserted through ports at opposite ends of the cavity. The system is connected to a Rohde \& Schwartz ZNB40 vector network analyzer (VNA) as shown in Figure \ref{fig:cavity}(c). This two-port configuration enables the measurement of the full scattering matrix (i.e., both reflection and transmission coefficients in amplitude and phase) as a function of driving frequency $\omega/2\pi$ and external magnetic field $\mu_0 H$. All measurements are done at room temperature.

In between posts, we placed two YIG spheres, securing them in place by mounting them in a 3D printed resin sample holder. The sample holder was specifically designed to accommodate up to four spheres, with the four positions located at the midpoints of each pair of adjacent posts. The sample holder's self-locking design onto the four posts ensures precise, repeatable positioning. The configurations considered in this work are shown in Figure \ref{fig:cavity}(d) and (e) which we will call the diagonal arrangement, DA and parallel arrangement, PA.

\begin{figure*}
    \centering
    \includegraphics[width=\linewidth]{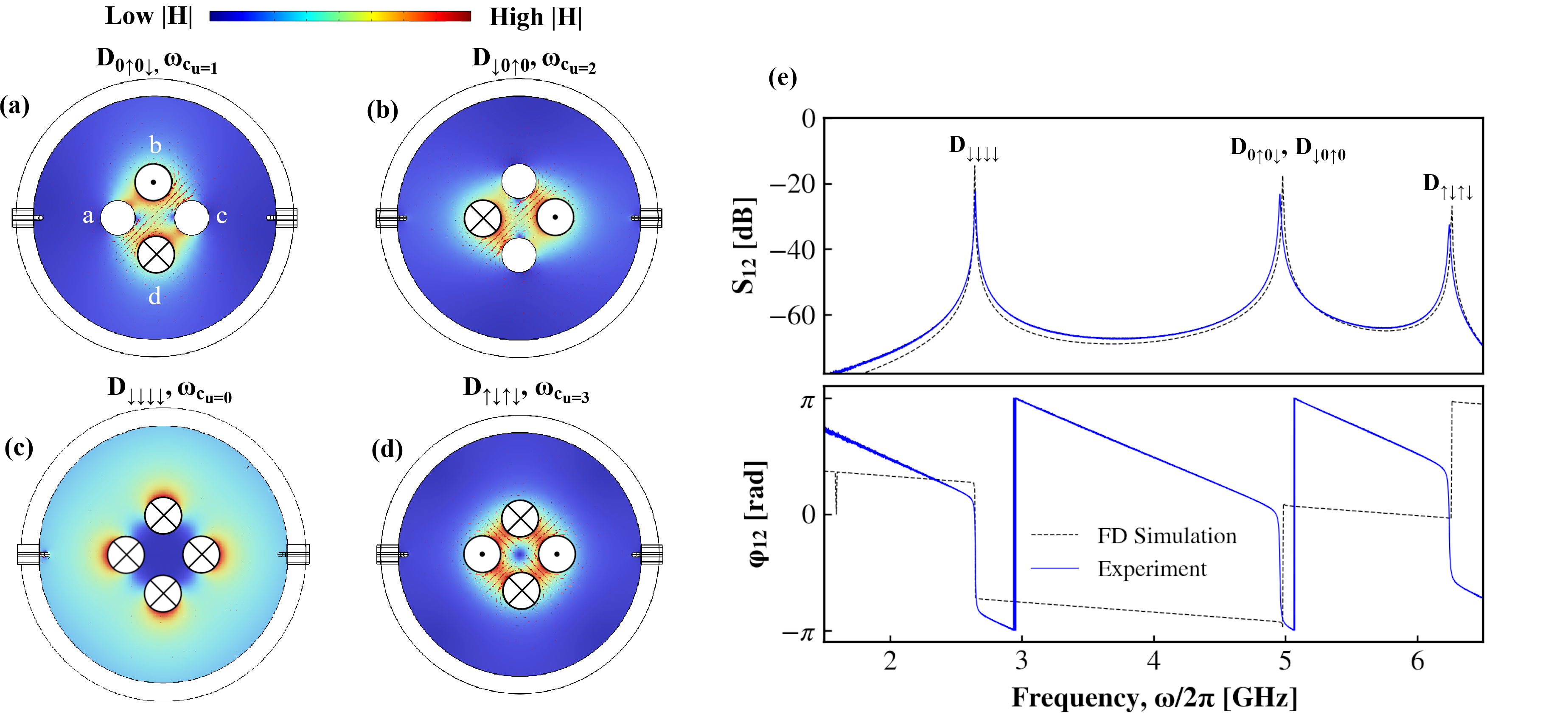}
    \caption{Eigenfrequency (EF) simulation of the MW magnetic field distribution inside the modeled empty cavity (a) $D_{0\uparrow0\downarrow}$, (b) $D_{\downarrow0\uparrow0}$, (c) $D_{\downarrow\downarrow\downarrow\downarrow}$ and (d) $D_{\uparrow\downarrow\uparrow\downarrow}$, the red (blue) regions indicate high (low) MW field intensity. The $\otimes$ and $\odot$ sign overlaid on the posts indicate out-of-post and into-post MW E-field direction, respectively. (e) Frequency sweep of the empty cavity with respect to the transmission parameter $S_{12}$ in dB and its transmission phase, $\varphi_{12}$ in rad. The blue solid line showing the transmission spectrum obtained from experiment, and the dashed black line is the same spectrum obtained from frequency domain (FD) simulation, applying perfect electrical conductor boundary condition at the inner walls of the cavity. The MW field distributions frequency peaks are labelled accordingly in (e). Note that the color scale of the MW magnetic field distribution are not normalized and are set to emphasize the MW field intensity distribution inside the cavity.}
    \label{fig:diagsweep}
\end{figure*}

Figure \ref{fig:diagsweep} presents the characterization of the empty cavity resonator. It supports electromagnetic modes that have spatially separated electric and magnetic components \cite{goryachevHighCooperativityCavityQED2014}. In effect, the MW electric field is concentrated in the gap, $d$, between the top of the posts and the inner lid surface of the cavity (see Figure \ref{fig:cavity}(b)). Simultaneously, the MW magnetic field distribution is concentrated in the area close to the posts as shown in the Figures \ref{fig:diagsweep}(a-d). We simulated the MW magnetic field distribution using eigenfrequency (EF) study within COMSOL Multiphysics\textsuperscript{\tiny\textregistered}' RF module. Figure \ref{fig:diagsweep}(a-d) shows the 2D MW magnetic field distributions at the plane where the spheres will be placed. The MW magnetic field focusing due to constructive interference of adjacent MW H-fields helps optimize the filling factor, $\eta$, which is expressed as  \cite{bourcinStrongUltrastrongCoherent2023, bourhillUniversalCharacterizationCavity2020, flowerExperimentalImplementationsCavitymagnon2019},

\begin{equation}
    \eta = \sqrt{\frac{\left(\int_{V_m} h_x \, \dv{V} \right)^2 + \left(\int_{V_m} h_y \, \dv{V} \right)^2}{V_m\int_{V_c} |h_x +h_y|^2\dv{V}}}.
    \label{eq:fillfactor}
\end{equation}

where $h_{x,y}$ are the $x$- and $y$-components of the cavity mode magnetic field, perpendicular to the z-axis that we set to coincide with $\mu_0 H$. $V_m$ and $V_c$ are the volume of the magnetic material and the cavity, respectively. By focusing the MW magnetic field within the sphere's volume we are able to achieve a strong coupling between the cavity photon and magnon modes according to the definition of coupling strength, $g_{uv}/2\pi$ \cite{bourcinStrongUltrastrongCoherent2023, bourhillUniversalCharacterizationCavity2020},

\begin{equation}
    \frac{g_{uv}}{2\pi} = \eta\sqrt{\omega_u} \frac{\gamma}{4 \pi} \sqrt{\frac{\mu}{g_L \mu_B} \mu_0\hbar n_s},
\end{equation}

where $u \left(v\right)$ indexes the internal cavity (magnon) modes, respectively taking values $u\in \{0, 1, 2, 3\}$ $(v \in \{ 0, 1 \})$. 

Shown in panel \ref{fig:diagsweep}(e) are the magnitude and phase of the transmission response. The solid blue and dashed lines represent the measured and simulated transmission responses, respectively. Three distinct resonance peaks are observed within the 1.5–6.5 GHz frequency range. We note the discrepancies between the simulated and the measured transmission phase. This discrepancy can be sourced to the electrical components of the VNA applying equipment-sourced electrical delay to the output.

Following the notation in Goryachev et al. \cite{goryachevCreatingTuneableMicrowave2015}, we use the notation, $D_{abcd}$, where $D$ denote the $D_4$ symmetry of the resonator posts, while $a,b,c,d$ denotes the post positions, starting from the leftmost post cycling clockwise. The posts can be assigned a 0, up, or down symbol depending on the $E_z$ direction at the post-to-lid gap, and so $a,b,c,d \in \{0, \uparrow, \downarrow\}$. 

One of the resonance peaks lies at frequency $\omega/2\pi=4.96$ GHz where we have degenerate peaks with an estimated total line width of $\gamma_{tot}/2\pi=21$ MHz. These peaks correspond to the $D_{0\uparrow0\downarrow}$ and $D_{\downarrow0\uparrow0}$ modes (Figures \ref{fig:diagsweep}(a) and (b)). This degeneracy arises due to the $D_4$ symmetry of the resonator posts, and that these modes resemble discrete whispering gallery modes \cite{goryachevCreatingTuneableMicrowave2015}. Specifically, the $D_{0\uparrow0\downarrow}$ mode involves a set of capacitive posts with out-of-post $E_z$ direction ($\odot$) and into-post ($\otimes$) $E_z$ direction shown in Figure \ref{fig:diagsweep}(a), while the $D_{\downarrow0\uparrow0}$ mode involves a corresponding set rotated by 90$^\circ$ shown in Figure \ref{fig:diagsweep}(b). Because the posts have nominally identical radii ($r=0.5$ mm) and post-to-lid gap ($d=130$ $\mu$m), these modes remain frequency-degenerate \cite{goryachevControllingWhisperinggallerydoubletmodeAvoided2014, kostylevSuperstrongCouplingMicrowave2016}. We also note the existence of lower-frequency $D_{\downarrow\downarrow\downarrow\downarrow}$ mode (2.65 GHz) and higher-frequency $D_{\uparrow\downarrow\uparrow\downarrow}$ mode (6.24 GHz) shown in Figures \ref{fig:diagsweep}(c) and (d). The $D_{\downarrow\downarrow\downarrow\downarrow}$ mode field is concentrated at the cavity periphery, whereas the $D_{\uparrow\downarrow\uparrow\downarrow}$ mode is concentrated at the interstitial regions between the posts. Note that the higher frequency modes beyond 6.24 GHz start at 17.5 GHz. Since it is located much farther from the strongly coupled modes at 5 GHz, we consider their effect to be negligible.

In multiple-post re-entrant cavities, the resonance frequency, $\omega_c / 2 \pi$ is only dependent on the geometry of the cavity which can be adjusted by modifying $d$. Because the resonance frequency is highly sensitive to $d$, machining tolerances inevitably introduce deviations from the nominal design. $d$ was therefore calibrated to 130.5 $\mu$m by matching simulated eigenfrequencies to the experimental spectrum \cite{bourhillUniversalCharacterizationCavity2020, depaulaExperimentsMatchSimulations2017} which was used for all subsequent FE simulations. Note that we apply perfect electric conductor (PEC) boundary condition at the inner cavity walls in the simulation to reduce simulation time.

Further in the text we will also refer to the cavity modes $D_{\downarrow\downarrow\downarrow\downarrow}$, $D_{0\uparrow0\downarrow}$ , $D_{\downarrow0\uparrow0}$, and $D_{\uparrow\downarrow\uparrow\downarrow}$ as 0th ($u=0$), 1st ($u=1$), 2nd ($u=2$), and 3rd ($u=3$) cavity modes, respectively. In the present cavity magnonics system, we treated two YIG spheres, positioned in two configurations shown in Figure \ref{fig:cavity}(d, e). We denote the resonant mode associated with the sphere in position 'YIG 1' as 0th ($v=0$) magnon mode and in position 'YIG 2' as 1st ($v=1$) magnon mode.

\section{\label{sec:theory}Theory}

\subsection{Internal and external phase}

The cavity-magnon effective internal coupling phase, $\theta_{uv}$, is the argument between the real $x$-component and the imaginary $y$-component of the complex MW cavity magnetic field that overlaps with the volume of the sphere. It is expressed by \cite{gardinManifestationCouplingPhase2023},

\begin{equation}
    \theta_{uv} = \text{arg}\left\{ \int_{V_{m_v}} \dv^3r \, h_u(r)\cdot \hat{x} + i \int_{V_{m_v}} \dv^3r \, h_u(r)\cdot \hat{y} \right\},
    \label{eq:phaseinternal}
\end{equation}

where $u$ and $v$ are the same internal mode indices discussed previously and $h_u$ is the MW magnetic field component of the $u$-th cavity mode. Note that the initial coordinate reference for $\theta_{uv}$ can be chosen arbitrarily. This choice does not affect observable physics, as only the relative phase shift between coupling pathways is physically relevant. Nevertheless, for clarity, we have illustrated our chosen coordinate reference in Figure 7 of the Appendix (also refer to Appendix E and the literature cited therein for more details on unitary transformation of the internal coupling phase).

It is possible to engineer $\theta_{uv}$ to specific values by carefully positioning the YIG spheres relative to the cavity field distributions. Here, we set P2 plane as the origin and increasing counter-clockwise. We then modify the coupling strength to include this phase term, i.e.,

\begin{equation}
    g_{uv} \Rightarrow g_{uv} e^{i\theta_{uv}}
\end{equation}

In addition to the internal coupling phase, another important parameter is the external coupling phase, $\phi_{up}$. This coupling phase depends on the parity of the MW magnetic field profile of the mode that is measured by the loop probe. The external phase has two possible values, $\phi_{up} = 0, \pi$, corresponding to an in phase or a $\pi$-delayed signal respectively which depend on the parity of the azimuthal MW magnetic field component sensed by the probe. Applying the first Markov approximation \cite{bourcinLevelAttractionQuasiclosed2024, wallsQuantumOptics1994, gardinerInputOutputDamped1985}, we assume the frequency independence of the cavity mode-bath coupling rate, and so the external dissipation rate becomes $ \kappa_{up} \left(\omega\right) \Rightarrow \kappa_{up} = \sqrt{\gamma_{up}} e^{i\phi_{up}}$, where $\gamma$ is real and represents the external cavity mode decay rate. $p \in \{0, 1\}$ indexes the bath probes P1 and P2.

Previous studies emphasize the importance of these coupling phases to get accurate frequency response of loop-coupled cavity magnonics system such as the works of Gardin et al \cite{gardinEngineeringSyntheticGauge2024, gardinManifestationCouplingPhase2023} and Bourcin et al \cite{bourcinLevelAttractionQuasiclosed2024}. In this system, coupling phases are critical for accurately describing the distinct spectral signatures associated with different YIG sphere configurations.

\subsection{S-parameters}

We model the system by considering a total Hamiltonian comprising 6 internal bosonic modes (4 cavity, 2 magnon modes) that are coupled to 2 bath modes. We write the total Hamiltonian as,

\begin{equation}
    H = H_{\text{int}} + H_{\text{ext}},
\end{equation}

where $H_{\text{int}}$ is the Hamiltonian terms of the internal bosonic modes following the second quantization formalism \cite{diracPrinciplesQuantumMechanics1981} and its cavity-magnon coupling quantified by $g_{uv}$. We define $H_{\text{int}}$ under the rotating wave approximation as 

\begin{equation}
    \begin{split}
        H_{\text{int}}/\hbar = \sum_{u=0}^{3}  &\tilde{\omega}_{c_u}\hat{c}_u^\dagger(\omega) \hat{c}_u(\omega)  + \sum_{v = 0}^1\tilde{\omega}_{m_v}\hat{m}_v^\dagger(\omega)\hat{m}_v(\omega)  \\ &+ \frac{1}{2} \sum_{u =0}^3 \sum_{v =0}^1  \left(g_{uv}\hat{c}_u(\omega) \hat{m}_v^\dagger(\omega)  + g_{vu}\hat{c}_u^\dagger(\omega) \hat{m}_v(\omega) \right),
    \end{split}
\end{equation}

where $\hat{c_u}$ $(\hat{c_u}^\dagger)$ the bosonic cavity mode annihilation (creation) operator of index $u\in\left[0, 3\right]$ obeying the commutation relation, $\left[c_u, c^\dagger_u\right] = 1$. Similarly, $m_v$ $(m_v^\dagger)$ is the magnon mode annihilation (creation) operator of index $v\in\{0, 1\}$ obeying $\left[m_v, m^\dagger_v\right] = 1$. Note that $c_u$ and $m_v$ commute with each other. The mode frequencies are then indicated by $\tilde{\omega}$ with index $c_u$ and $m_v$ corresponding to the cavity and magnon modes respectively where $\tilde{\omega}_{c_u(m_v)} = \omega_{c_u(m_v)} -i \kappa_{c_u(m_v)}/2$ and $\kappa$ is the total dissipation rate of the mode with the same index.

\begin{figure*}
    \centering
    \includegraphics[width=\linewidth]{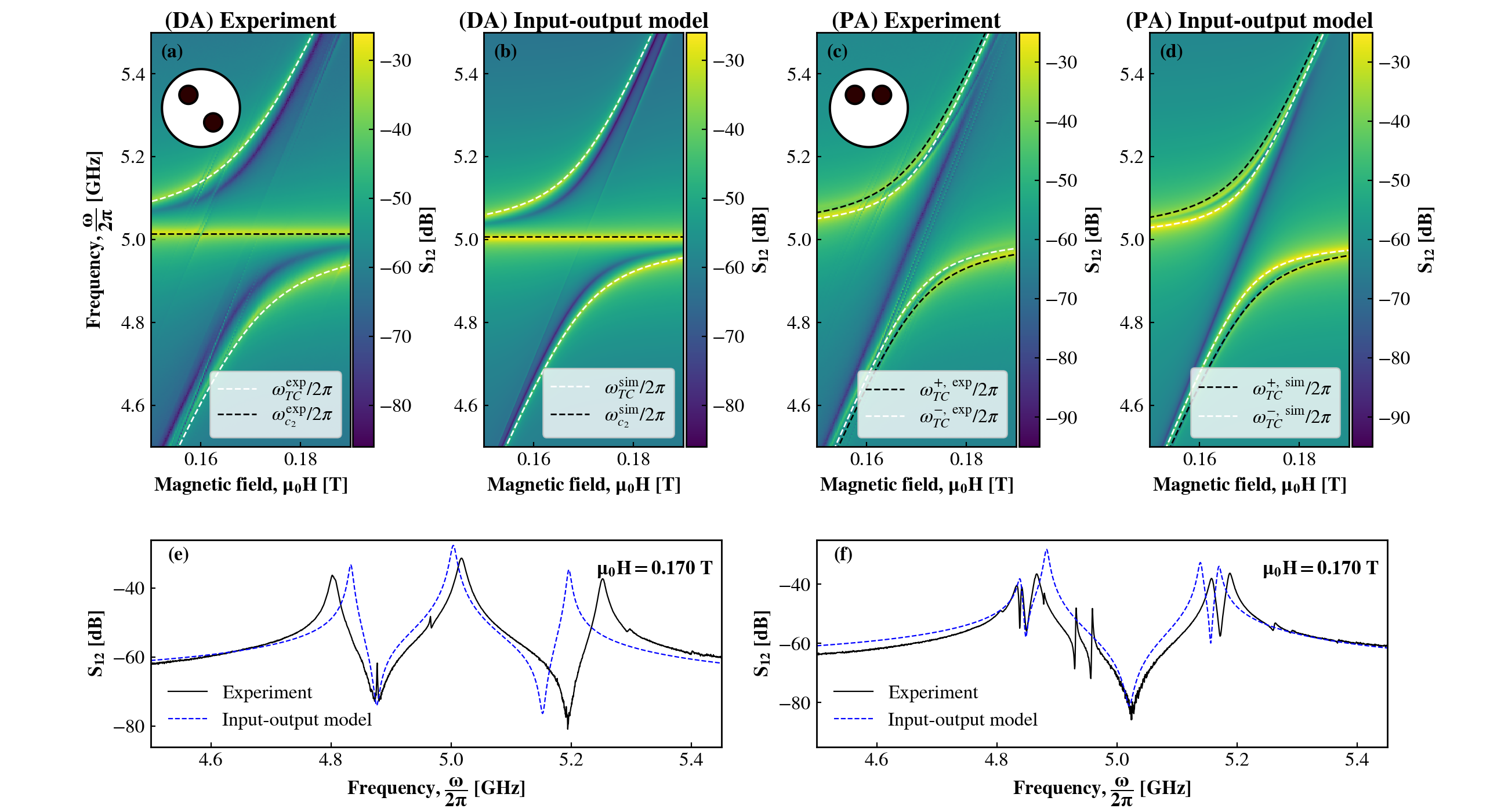}
    \caption{ (a-d) Scattering parameter $S_{12}$ color map in dB scale as a function of frequency, $\omega/2\pi$ and external magnetic field, $\mu_0 H$ for the DA and PA. Panel (a) and (b) are the measured and calculated response for the DA configuration, respectively. The dashed lines in panel (a-d) are the eigenfrequencies of the polariton, $\omega_{TC}$, modelled using the two-level Tavis-Cummings coupling model (see Appendix \ref{app:Tavis} and the relevant literature \cite{bourcinSpincavitronicsRepulsiveAttractive2024, tavisExactSolution$N$MoleculeRadiationField1968} cited therein). Panel (c) and (d) on the other hand, correspond to the PA. To obtain the numerically calculated response in (b) and (d), we feed parameter values obtained through coupled Landau-Lifshitz-Gilbert and Maxwell's equation FE simulation (listed in Table \ref{tab:calc_params} in the Appendix \ref{app:comsol}) to Eq. (\ref{eq:sparams}). The diagram on the top left of subfigure (a, c) depicts the position of the YIG inside the cavity. (e, f) Frequency line cuts at $\mu_0 H = 0.170$ T for DA and PA, respectively. The black solid line corresponds to measurement, while the blue dashed line corresponds to numerical calculation.}
    \label{fig:trans}
\end{figure*}

Next, we have the harmonic oscillator bath modes describing the probes and its interaction with the cavity modes expressed by the external Hamiltonian,

\begin{equation}
\begin{split}
        H_{\text{ext}}/\hbar = \int_\mathbb{R} &\dv\omega \sum_{p=0}^1  \, b^\dagger_p(\omega) b_p(\omega)  \\ & +\frac{i}{\sqrt{2 \pi}} \int_\mathbb{R} \dv\omega \sum_{u  = 0}^3 \sum_{p=0}^1 \, (\kappa_{up} b^\dagger_{p}(\omega) c_u(\omega)  - h.c),
\end{split}
\end{equation}

where $\kappa_{up}$ is the coupling rate of the cavity mode to the bath probes with index $p \in \{0, 1\}$ corresponding to probes P1 and P2 respectively.

Based on the defined Hamiltonian, the $2 \times 2$ S-parameter matrix, $\textbf{S}$ for the cavity modes can be obtained from standard input-output formalism for degenerate modes of a cavity with two ports \cite{zhangBroadbandNonreciprocityEnabled2020, gardinerInputOutputDamped1985}. The  S-matrix in the frequency domain is expressed as,

\begin{equation}
\label{eq:sparams}
    \textbf{S} = \textbf{C} + \textbf{D} \, \left[ -i{\omega} \textbf{I} - \textbf{A} \right]^{-1} \,\textbf{B}
\end{equation}

where we have matrix $\textbf{A}$ that corresponds to the internal modes and their interactions, matrix $\textbf{B}$ that corresponds to the interaction of the cavity mode with the bath probes and crosstalk matrix, $\textbf{C}$ with $0 \leq \xi \leq 1$ representing the cross talk between the ports. Lastly, $\textbf{I}$ is a $6\times6$ identity matrix.

\begin{equation}
    \begin{split}
        \textbf{A} &= \begin{pmatrix}
            -i\tilde\omega_{c_0} & 0 & 0 & 0 & i\tilde{g}_{00} & i\tilde{g}_{01} \\
            0 & -i\tilde\omega_{c_1} & 0 &0 & i\tilde{g}_{10} & i\tilde{g}_{11}\\
            0 &0 & -i\tilde\omega_{c_2} & 0 & i\tilde{g}_{20} & i\tilde{g}_{21} \\
           0 & 0 & 0 & -i\tilde\omega_{c_3} & i\tilde{g}_{30} & i\tilde{g}_{31} \\ 
           i\tilde{g}_{00}^* & i\tilde{g}_{10}^* & i\tilde{g}_{20}^* & i\tilde{g}_{30}^* & -i\tilde\omega_{m_0} & 0 \\ 
           i\tilde{g}_{01}^* & i\tilde{g}_{11}^* & i\tilde{g}_{21}^* & i\tilde{g}_{31}^* & 0 & -i\tilde\omega_{m_1}
        \end{pmatrix},\\
        \textbf{B}^\text{T} &= \begin{pmatrix}
            \sqrt{ \gamma_{00} }e^{i \phi_{00}}  & \sqrt{ \gamma_{10} }e^{i \phi_{10}} & \sqrt{ \gamma_{20} }e^{i \phi_{20}} & \sqrt{ \gamma_{30} }e^{i \phi_{30}} & 0 & 0 \\
            \sqrt{\gamma_{01} }e^{i \phi_{01}} & \sqrt{ \gamma_{11} }e^{i \phi_{11}} & \sqrt{ \gamma_{21} }e^{i \phi_{21}} & \sqrt{ \gamma_{31} }e^{i \phi_{31}} & 0 & 0 \\
        \end{pmatrix},\\
        \textbf{C} &= \begin{pmatrix}
            \sqrt{1-\xi} & \sqrt{\xi} \\
            \sqrt{\xi} & \sqrt{1 - \xi}
        \end{pmatrix}\\
        \textbf{D} &= -\mathbf{CB}^\dagger, \\
    \end{split}
    \label{eq:smatrices}
\end{equation}

since we assume no crosstalk between the ports, we set $\xi = 0$. Note that we write $\tilde{g}_{uv} = g_{uv} e^{i\theta_{uv}}$ with $\tilde{g}_{uv}^*$ as its complex conjugate, for compactness. In reproducing the experimental observation, we extracted the parameters $\omega_{c_u}, g_{uv}, \theta_{uv}$ and $\phi_{up}$ values from FE simulation. We use results from both EF and FD solvers (detailed in Appendix \ref{app:comsol}) and use it as input to numerically solve the S-parameter matrix, $\textbf{S}$.

\section{\label{sec:results}Results}

\subsection{\label{subsec:transres}Cavity transmission response}

We investigated the two configurations mentioned previously in Section \ref{subsec:fourpost}, namely the DA and PA (diagram of the configurations are shown in Figure \ref{fig:cavity}(d, e) and are also depicted within subfigures \ref{fig:trans}(a,c)). The backward transmission response, $S_{12}$ of both configurations are shown in Figure \ref{fig:trans}, with (a, c) showing the experimental response and (b, d) showing the calculated response based on Eq. (\ref{eq:sparams}). The data measurements were taken within the frequency range $4.00 < \frac{\omega}{2 \pi} < 7.25$ GHz, with frequency step size $\Delta f = 5$ MHz, and external field $0.114 < \mu_0 H < 0.248$ T, with external field step size, $0.25$ mT. We set the VNA intermediate frequency bandwidth (IFBW) at $1.0$ kHz and source power at $-10$ dBm.

Dependent on the YIG spheres' position, the measured backward transmission, $S_{12}$ of both configurations show distinct signatures. When the two YIG spheres are positioned in the DA, we observe three resonance peaks at a magnetic field value, $\mu_0 H = 0.170$ T as shown in Figure \ref{fig:trans}(e) with the black solid line representing the measurement and blue dashed line representing numerically calculated input-output model. One peak corresponds to an uncoupled mode, observed at the frequency, $\omega/2\pi = 5.03$ GHz. On the other hand, the two peaks corresponding to the CMP, exhibit strong coupling with effective coupling strength, $g_{DA}^{\text{exp}}/2\pi = 160$ MHz, which is larger than the resonant mode line width. In Figure \ref{fig:trans}(a, b), the uncoupled mode resonance frequencies, $\omega_{c_2}/2\pi$ are indicated by the dashed black line, while the eigenfrequencies obtained using the Tavis-Cummings model (see Appendix \ref{app:Tavis}) for the coupled mode, $\omega_{TC}/2\pi$ are indicated by the dashed white line. On the contrary, placing the YIG spheres in the PA leads to four resonance peaks, which correspond to both modes being coupled to the YIG spheres with coupling strengths $g_{PA}^{-, \text{ exp}}/2\pi = 106$ MHz and $g_{PA}^{+, \text{ exp}}/2\pi = 127$ MHz. In Figure \ref{fig:trans} legends, the $\pm$ sign indicates higher and lower coupling strength respectively, and the superscripts "$\text{exp}$" and "$\text{sim}$" indicate value obtained from experiment or simulation respectively. These four resonances can be seen in Figure \ref{fig:trans}(f). We note some observations of narrow peaks owing to higher order however we consider interaction with these modes to be out of scope of our current work.

In Figure \ref{fig:trans}(b) and (d) we calculated the $S_{12}$ response using Eq. (\ref{eq:sparams}). The parameter values used in the calculation are obtained from both FE simulation (coupling strengths, coupling phases, resonance frequency) and measurements (effective gyromagnetic ratio, magnon mode linewidth, cavity mode linewidth). These values are listed in the Appendix \ref{app:comsol} under Table \ref{tab:calc_params}. We do not exclusively use simulated parameter values obtained from one type of FE study (only FD or only EM). This is because some values are better estimated in FD where we include the probe drive, such as the coupling strength, $g_{uv}$ and internal phase $\theta_{uv}$. The probe drive in this case produces the system's response to excitation that is dependent on how the probes are placed in the cavity. On the other hand, results from the EF study better estimate parameters such as the decay rate and external phase, as the FD study is performed with an active excitation port, which obscures the extraction of these parameters. Note that we perform the FD simulation with $\mu_0H$ set to be much greater than the YIG saturation magnetic field and that the FMR is far off resonance with the cavity mode. Extraction details of simulated parameters are described in Appendix \ref{app:comsol}.

Using the input-output model extended with internal and external coupling phases, we are able to reproduce the measured response main features, namely the uncoupled mode in the DA, and the four peaks resonances apparent in the PA. Although we are able to get good agreement of the main features compared with measurement results, we note that the coupling strength value obtained from simulation is slightly underestimated, with effective coupling strength $g_{DA}^{sim}/2\pi = 128$ MHz, $g_{PA}^{+, \text{ sim}}/2\pi = 121$ MHz and $g_{PA}^{-, \text{ sim}}/2\pi = 92$ MHz, corresponding to offsets of 21.45\%, 4.83\%, and 11.28\%. This discrepancy may be traced to the simulation not perfectly reflecting real measurement conditions. For example, a positional mismatch between measurement and simulation of the YIG spheres relative to the MW magnetic field distribution could lead to a mismatch in the filling factor, $\eta$ and thus the coupling strength, while mismatch between the post-to-lid gap lead to a mismatch in the resonance frequency of modes. The mismatch in the antiresonance frequency might also be affected by the existence of higher magnetostatic modes which is not considered in the model.

Despite the discrepancies in the effective coupling strength and frequency shift between measurement and simulation, the matrix formalism derived from the input-output theory captures the essential physics. Specifically, the $6 \times6$ matrix \textbf{A} is sufficient to reproduce the distinct transmission signatures of both configurations: the three-peak response in the DA and the four-peak polariton response in the PA. This is because the full $6 \times 6$ matrix \textbf{A} encodes all pairwise couplings, resonance frequencies, and internal coupling phases of the modes that is critical for the manifestation of the interference effects. Although the full matrix is a more complete description, expanding it for an analytical expression, is impractical. This impracticality motivates a reduced description of the system, i.e. isolating the interference mechanism responsible for these transmission signatures to help us unravel the distinct signature of DA and PA. The reduced $4 \times4$ matrix containing only the two degenerate cavity modes, $\omega_{c1}$ and $\omega_{c2}$ and two magnon modes $\omega_{m0}$ and $\omega_{m1}$ is written as,

\begin{equation}
    \textbf{A}_{\text{red}} = \begin{pmatrix}
        -i\tilde{\omega}_{c1} & 0 & ig_{10}& ig_{11}\\
         0 & -i\tilde{\omega}_{c2} & ig_{20} & ig_{21}e^{i\theta_p} \\
        ig_{10} & ig_{20} & -i\tilde{\omega}_{m0} & 0 \\
        ig_{11}& ig_{21}e^{-i\theta_p} & 0 & -i\tilde{\omega}_{m1}
    \end{pmatrix},
    \label{eq:reduced}
\end{equation}

where the internal coupling phases accumulate as the physical phase, $\theta_p = \theta_{21} -\theta_{11} - (\theta_{20} -\theta_{10})$, after applying a unitary transformation (see Appendix \ref{app:unitrans}). The unitary transformation also reveals internal-external coupling phase interference in the cavity-bath interaction term which for a two-probe cavity system is represented by a reduced matrix \textbf{B},

\begin{equation}
    \textbf{B}_{\text{red}} = \begin{pmatrix}
        \sqrt{\gamma_{10}}e^{i(\phi_{10} - \theta_{10})} & \sqrt{\gamma_{11}}e^{i(\phi_{11} - \theta_{10})}\\
         \sqrt{\gamma_{20}}e^{i(\phi_{20} - \theta_{20})} & \sqrt{\gamma_{21}}e^{i(\phi_{21}  - \theta_{20})}  \\
        0 & 0  \\
        0 & 0
    \end{pmatrix},
    \label{eq:reducedB}
\end{equation}

Solving Eq. (\ref{eq:smatrices}) for matrices (\ref{eq:reduced}) and (\ref{eq:reducedB}) and the fact that $\phi_{10, 11,20,21} = 0$ from Table \ref{tab:calc_params} in Appendix \ref{app:comsol}, we can obtain a reduced analytical $S_{21}$ expression, 

\begin{widetext}
\begin{equation}
\begin{aligned}
S_{21} = &\frac{i}{D}\Biggl(\sqrt{\gamma_{10}\gamma_{11}} (\Delta_{c_2}\Delta_{m_0}\Delta_{m_1} - \Delta_{m_0}g_{21}^2 - \Delta_{m_1}g_{20}^2) + \sqrt{\gamma_{10}\gamma_{21}} (\Delta_{m_0}g_{11}g_{21}e^{i\theta_p} + \Delta_{m_1}g_{10}g_{20})e^{i(\theta_{20} - \theta_{10}) }  \\
&+\sqrt{\gamma_{20}\gamma_{11}} (\Delta_{m_0}g_{11}g_{21}e^{-i\theta_p} + \Delta_{m_1}g_{10}g_{20})e^{-i( \theta_{20} - \theta_{10} )}  + \sqrt{\gamma_{20}\gamma_{21}} (\Delta_{c_1}\Delta_{m_0}\Delta_{m_1}- \Delta_{m_0}g_{11}^2 - \Delta_{m_1}g_{10}^2)\Biggr),
\end{aligned}
\label{eq:s21}
\end{equation}

\begin{equation}
\begin{aligned}
D = \Delta_{c_1}\Delta_{c_2}\Delta_{m_0}\Delta_{m_1} - \Delta_{c_1}\Delta_{m_0}g_{21}^2 - \Delta_{c_1}&\Delta_{m_1}g_{20}^2 - \Delta_{c_2}\Delta_{m_0}g_{11}^2 \\&- \Delta_{c_2}\Delta_{m_1}g_{10}^2 + g_{10}^2 g_{21}^2 - 2g_{10}g_{11}g_{20}g_{21}\cos{(\theta_p)} + g_{11}^2 g_{20}^2,
\end{aligned}
\end{equation}
\end{widetext}

where, $\Delta_{c_u(m_v)} = \omega- \tilde{\omega}_{c_u(m_v)}$.

\begin{figure}
    \centering
    \includegraphics[width=\linewidth]{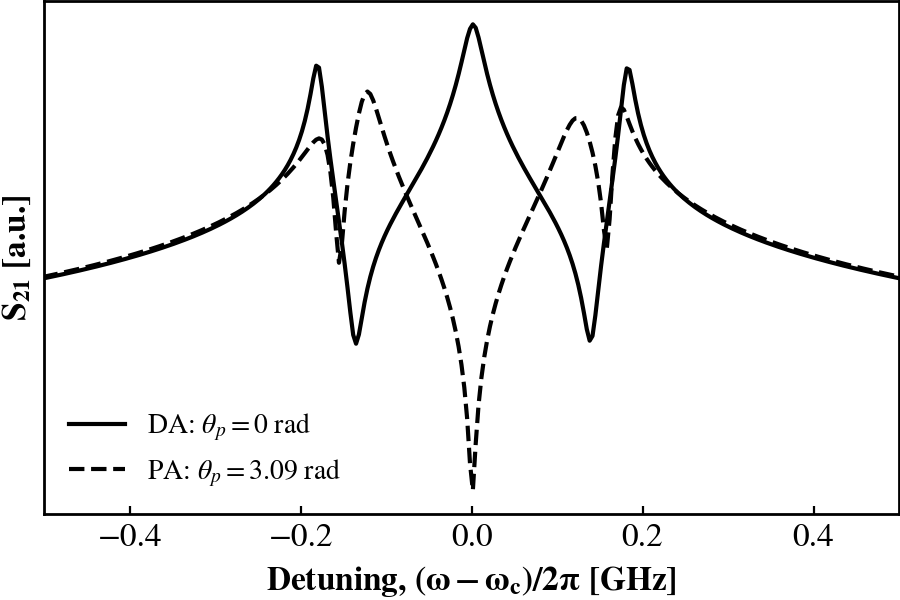}
    \caption{Numerical calculation of Eq. (\ref{eq:s21}) when the physical phase, $\theta_p = 0$ rad corresponding to three peaks response and the uncoupled mode (solid black line) and $\theta_p = 3.09$ rad corresponding to four peaks of CMPs (dashed black line). The physical phase significantly affects the transmission signature of a multimode cavity magnonics system.}
    \label{fig:analy}
\end{figure}

The reduced analytical expression simplifies $S_{21}$ and allows find solutions for the resonance and antiresonance eigenfrequencies. In the case of DA, substituting the values in Table \ref{tab:calc_params}, we have $\theta_p = 0$ rad, this leads to the signature three peaks transmission response in Figure \ref{fig:analy} (solid black line). On the other hand, for PA, as we have $\theta_p = 3.09$ rad the transmission response shows two pairs of polariton branches, coupled asymmetrically with no presence of the uncoupled mode (dashed black line). Using Eq. \ref{eq:s21}, we successfully capture the distinct transmission signatures of both the DA and PA configurations, in agreement with the full $6 \times 6$ matrix. This shows that the main modes are sufficient to explain the distinct signature of PA and DA and that we can continue our analysis for three cases, when (1) $\theta_p = 0$ rad and when (2) $\theta_p = \pi$ rad and finally attempt a more general case of (3) arbitrary $\theta_p$.

\subsubsection{Case 1: $\theta_p = 0$ rad}

When $\theta_p = 0$ rad, the coupling pathways interfere, resulting in three resonant peaks. This can be shown by finding solutions for the resonances when the $S_{21}$ denominator D tends to $0$, $D \rightarrow0$. To simplify the derivation, we assume the degeneracy of cavity modes and similarly, the magnon modes. We also assume equal coupling between the modes. These assumptions are approximate and comes from the fact that the resonance frequency of the degenerate cavity modes, the magnon modes, and the coupling strengths are close in value ($\omega_{c_1} \approx \omega_{c_2}$, $\omega_{m_1} \approx \omega_{m_2} = \gamma \mu_0 H$, and $g_{u1} \approx g_{u2}$, respectively, also see Table \ref{tab:calc_params}). This equality can be written as,

\begin{equation}
\begin{aligned}
       \Delta_{c_1} &= \Delta_{c_2}=\Delta_c  \\
     \Delta_{m_0} &= \Delta_{m_1} =\Delta_m  \\
    g_{10} &= g_{11} = g_{20} = g_{21} = g.
\end{aligned}
\label{eq:condition}
\end{equation}

Substituting Eq. (\ref{eq:condition}) to Eq. (\ref{eq:s21}) we get three resonant conditions for the case $\theta_p = 0$. First, we have a pure magnon mode resonance corresponding to a solution where $\Delta_m = 0$. Second, a solution where $\Delta_c = 0$. This corresponds to a resonance that is invisible to the magnon modes (the uncoupled mode). And lastly, when $\Delta_c\Delta_m = 4g^2$, we have a coupled cavity-magnon mode. In the transmission response, only resonances corresponding to the cavity photons, $\Delta_c = 0$ and $\Delta_c\Delta_m = 4g^2$ will be observed while pure magnonic resonances will be dark to the probe. By minimizing $D$, we find that the CMP has eigenfrequencies,

\begin{equation}
    \omega_\pm|_{\theta_p = 0} = \frac{1}{2}\left( \omega_c + \omega_m \pm \sqrt{(\omega_c-\omega_m)^2+4g_{\text{eff}}^2}\right),
    \label{eq:polariton0}
\end{equation}

with an effective coupling strength, 

\begin{equation}
    g_{\text{eff}}^2|_{\theta_p = 0} = 4g^2,
\end{equation}

assuming the coupling strength has equal values. This confirms the observation in the transmission response of DA: A pair of CMP and an uncoupled mode. Notably, this response resembles previously reported chiral circulating CMP in \cite{yuCirculatingCavityMagnon2020, bourhillGenerationCirculatingCavity2023}. Unlike the chiral coupling due to circularly polarized MW field, the transmission response we observed were due to coupling pathway interferences accumulated in the physical phase, $\theta_p$, effectively cancelling each other. 

Further, we can determine the antiresonance frequencies by minimizing the numerator of Eq. (\ref{eq:s21}). This yields,

\begin{equation}
    \omega_\pm^{\text{ar}}|_{\theta_p = 0} = \frac{1}{2}\left( \omega_c+\omega_m \pm \sqrt{(\omega_c-\omega_m)^2+4g_\text{ar, eff}^2}\right),
\end{equation}

where $X_1 = \sqrt{\gamma_{10}\gamma_{11}} + \sqrt{\gamma_{20}\gamma_{21}}$ is the phase-independent transfer terms and $X_2^{(0)} = \sqrt{\gamma_{10}\gamma_{21}}e^{i(\theta_{20} - \theta_{10})} + \sqrt{\gamma_{20}\gamma_{11}}e^{-i( \theta_{20} - \theta_{10} )}$ define the phase-dependent crosstalk between probes for $\theta_p=0$ case. The antiresonance also exhibits an avoided crossing with the FMR mode, characterized by, 

\begin{equation}
    g_\text{ar, eff}^2|_{\theta_p = 0} = 2g^2\left(\frac{X_2^{(0)}}{X_1}- 1\right).
    \label{eq:antires_0}
\end{equation}

\subsubsection{Case 2: $\theta_p = \pi$ rad}

Following the same procedure as the previous case, the derivation leads to the solution,

\begin{equation}
    \Delta_c\Delta_m = 2g^2.
    \label{eq:numerator}
\end{equation}

This solution corresponds to CMP eigenfrequencies,

\begin{equation}
    \omega_\pm|_{\theta_p=\pi} = \frac{1}{2}\left( \omega_c + \omega_m \pm \sqrt{(\omega_c-\omega_m)^2+4g_{\text{eff}}^2}\right).
    \label{eq:polaritonpi}
\end{equation}

where the effective coupling strength differs, 

\begin{equation}
    g_\text{eff}^2|_{\theta_{p} = \pi} = 2g^2.
\end{equation}

In the $\theta_{p} = \pi$ rad case, the effective coupling strength will be smaller compared to the $\theta_{p} = 0$ case. This effect is consistent to measurements. The antiresonance on the other hand has frequencies corresponding to the FMR,

\begin{equation}
    \omega_{\text{ar}}|_{\theta_p =\pi} = \omega_m.
\end{equation}

Although there is a solution where coupled antiresonances exist, this will not be seen in an ideal $\theta_p =\pi$ rad case as the transmission response $S_{21}$ pole has a second-order form, overlapping and overpowering the $S_{21}$ zero,

\begin{equation}
    S_{21} \propto \frac{\Delta_m \left(\Delta_c\Delta_m -2g^2\right)}{\left(\Delta_c\Delta_m -2g^2\right)^2} = \frac{\Delta_m}{\Delta_c\Delta_m -2g^2} .
\end{equation}

\subsubsection{Case 3: Arbitrary $\theta_p$}

In the simulation, we do not obtain an exact $\theta_p = 0, \pi$ rad but rather $\theta_p = 3.09$ rad and so solutions for an arbitrary phase is desirable. We will see that, a value other than $\theta_p = 0, \pi$ rad will lift the degeneracy of the CMPs. The physical phase, in this case, explains the distinct two dips in between the pairs of resonance peaks in PA. This can be demonstrated by expressing the solutions for an arbitrary phase, which leads to the solutions of enhanced (superscript $+$) and attenuated (superscript $-$) polariton pairs,

\begin{equation}
    \omega_\pm^{(\pm)}|_{\theta_p}= \frac{1}{2} \left(\omega_c + \omega_m \pm \sqrt{(\omega_c-\omega_m)^2 + 4g_{\text{eff}}^{(\pm)}(\theta_p)}\right)
\end{equation}

where, $g_{\text{eff}}^{(\pm)} (\theta_p) = 2g^2 \pm \sqrt{2g^4\left[1+\cos{(\theta_p)}\right]}$ governs how the coupling strength varies due to internal coupling phase interferences. In addition to the antiresonance frequency $\omega_{ar} = \omega_m$, we also have a coupled antiresonance that can be expressed like Eq. (\ref{eq:antires_0}) with a $\theta_p$-shifted, phase-dependent crosstalk term,

\begin{equation}
    g_\text{ar, eff}^2|_{\theta_p } = 2g^2\left(\frac{X_2^{(0)} + X_2^{(\theta_p)}}{X_1}- 1\right).
    \label{eq:antires_arb}
\end{equation}

where we have $X_2^{(\theta_p)} = \sqrt{\gamma_{10}\gamma_{21}}e^{i( \theta_{20} - \theta_{10} +\theta_p)} + \sqrt{\gamma_{20}\gamma_{11}}e^{-i( \theta_{20} - \theta_{10} - \theta_p)}$.

To conclude, we interpret these distinct signatures resulting from different YIG sphere configurations as manifestations of interfering internal coupling phases. The coupling between cavity modes and magnon modes are complex, with the imaginary part containing the phase factor, which governs the interference mechanism. Internal coupling phases are usually trivial in cases involving a single magnon interacting with multiple cavity modes. This can be shown by applying unitary transformation (see Appendix \ref{app:unitrans}) to the Hamiltonian which leads to the vanishing of the coupling phase (see Gardin et al. \cite{gardinEngineeringSyntheticGauge2024, gardinManifestationCouplingPhase2023}). Unlike the systems previously mentioned, we can consider our system as a loop-coupled multimode system \cite{yuanLoopTheoryInputoutput2020} where unitary transformation leads to an accumulation of these internal phase that emerge as a physical phase. This physical phase manifests in the response of the system, resulting in the distinct DA and PA response. It is also worth noting that the manifestation of the same interference in the dispersive regime has been studied by Gardin et al. \cite{gardinManifestationCouplingPhase2023} where the polaritons corresponding to different modes can either show repulsion or cross each other, the latter indicating no interaction between them. Experimentally, one can modify the phase relationship between internal coupling pathways, $\theta_p$, simply by moving the YIG spheres to some special positions. Doing so will cause the transmission signature to change, such when switching between DA and PA. In the general case, when $\theta_p \neq 0$ rad we will see two pairs of CMPs as evidenced in the transmission response of PA in Figure \ref{fig:trans}. On the other hand, when $\theta_p = 0$ rad three solutions exist showing the chiral-like transmission corresponding to the degenerate modes. Although this reduced model is able to predict the transmission signature, explaining nonreciprocity at the antiresonance requires the inclusion of $\pi$-delayed far-detuned modes into the model.

\begin{figure*}
    \centering
        \includegraphics[width=0.90\linewidth]{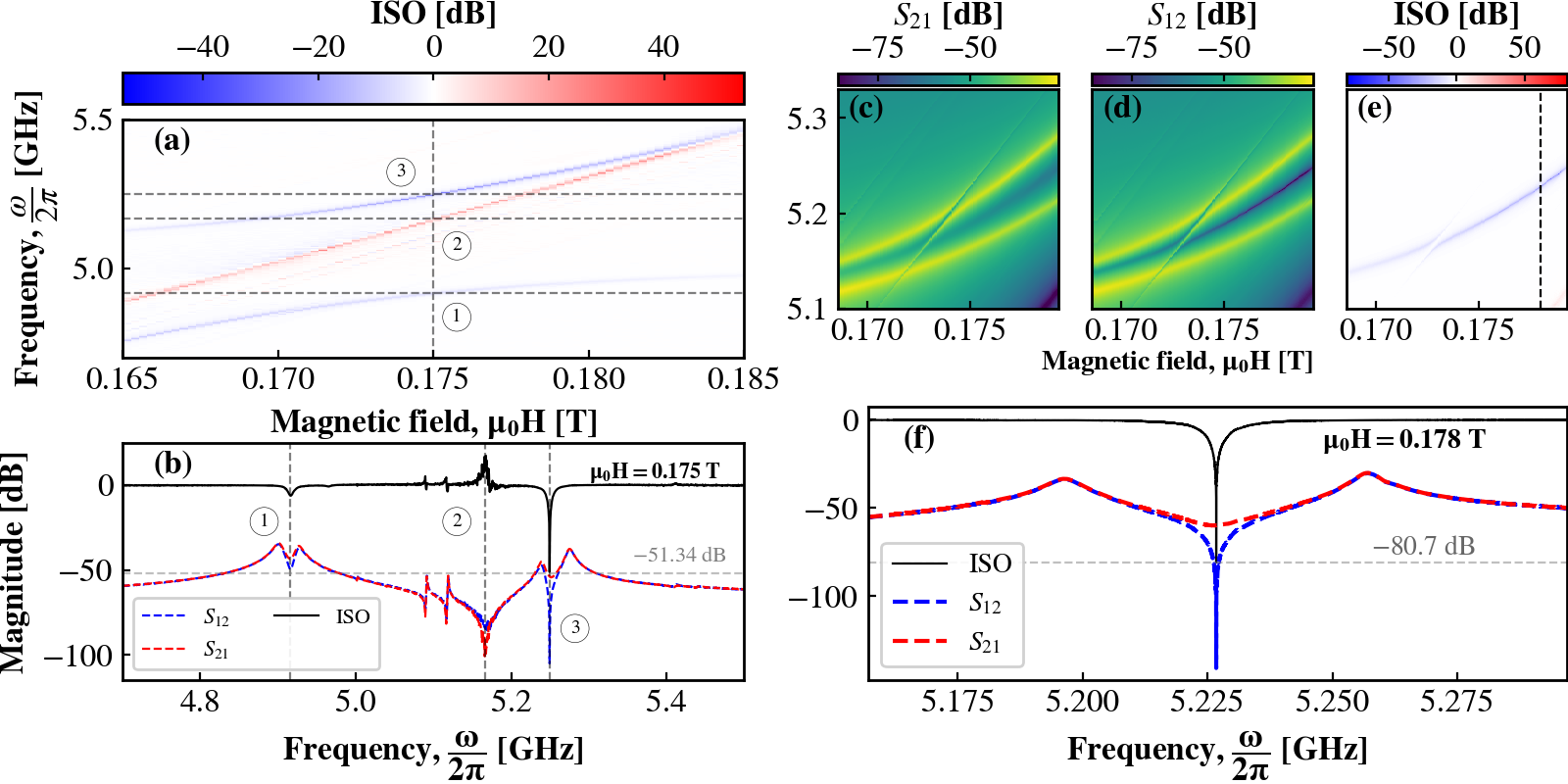}
    \caption{(a) Overview of the nonreciprocity quantified by the isolation ratio, ISO. Three branches are apparent which corresponds to the antiresonances, with red indicating larger $S_{12}$ and blue indicating larger $S_{21}$. The branches are apparent at (1) between the lower polariton resonances, (2) at the FMR frequency, and (3) between the two upper polariton resonances. The nonreciprocity occurs at the antiresonance frequencies, shown in the frequency sweep at $\mu_0 H=0.175$ T (b), where we observe peak ISO of -51.34 dB corresponding to the upper polariton frequency. The gray lines are visual guides. (c-e) A closer look at the antiresonance between the upper polariton resonances, in this window we observed nonreciprocity of up to -80.7 dB at $\mu_0 H=0.178$ T. (f) Frequency line cut at $\mu_0 H=0.178$ T}
    \label{fig:nonrecip}
\end{figure*}

\subsection{\label{nonreciprocal}Nonreciprocal transmission at antiresonance}
Notably, nonreciprocal transmission was observed at antiresonance frequencies in the PA configuration. We can quantify this nonreciprocity through the isolation ratio, ISO, that relates to the non-equal transmission between the forward and backward transmission directions, and is expressed by,

\begin{equation}
    \text{ISO [dB]}=S_{12} \text{ [dB]} - S_{21} \text{ [dB]}.
    \label{eq:iso}
\end{equation}

In this equation, ISO can be positive or negative. The sign indicates the direction of the nonreciprocal transmission.

An overview of the ISO response is plotted in Figure \ref{fig:nonrecip}(a), plotted from the same data used to plot Figure \ref{fig:trans}(a, b). In this cavity magnonics system, we identified three ISO branches. These branches lie at the antiresonance frequencies of the lower polariton branch, the FMR, and upper polariton branch labelled (1), (2), and (3) respectively. The corresponding nonreciprocal transmission at $\mu_0 H = 0.175$ T is plotted in Figure \ref{fig:nonrecip}(b), where the crossing points from \ref{fig:nonrecip}(a) are marked accordingly with the same labels alongside the gray dashed line that serves as a visual guide. At the mentioned external field value, we observed nonreciprocity of up to -51.34 dB. This observation motivates us to measure in the region around the high value of nonreciprocity with better resolution.

To better resolve the frequency around the nonreciprocity we refined three measurement parameters: narrowing the step size to $15$ kHz over the $5.1 - 5.33$ GHz frequency window, decreasing the external field step size to 0.04 mT within an external field window $0.1687 < \mu_0 H < 0.1792$, and reducing the VNA intermediate frequency (IF) bandwidth to $0.5$ kHz to minimize the noise floor with source power set to $0$ dBm. A comparison between $S_{21}$ and $S_{12}$ shown in Figure \ref{fig:nonrecip}(c) and (d) respectively, shows nonreciprocal transmission at the antiresonance (shown in Figure \ref{fig:nonrecip}(e)), located between two resonances of the upper branch polariton. A frequency line cut at $\mu_0 H = 0.178$ T shows an ISO value reaching -80.7 dB, suggesting a large phase-dependent nonreciprocity.

Although the reduced matrix, $\textbf{A}_{\text{red}}$ were able to explain the origin of the distinct DA/PA signatures, the description fails to explain the nonreciprocal transmission we see in the experiment. Thus, to reproduce this nonreciprocity using the model, we incorporated lower frequency $D_{\downarrow\downarrow\downarrow\downarrow}$ mode, and higher frequency $D_{\uparrow\downarrow\uparrow\downarrow}$ in the matrix $\textbf{A}$ and $\textbf{B}$ of Eq. (\ref{eq:smatrices}). The addition of these modes to the input-output model, takes into account the local effect of far-detuned modes to the antiresonance close to the coupled $D_{0\uparrow0\downarrow} \text{ and } D_{\downarrow0\uparrow0}$ modes. The effect, mediated by the coupling phases of $D_{\downarrow\downarrow\downarrow\downarrow}$ and $D_{\uparrow\downarrow\uparrow\downarrow}$ mode induces nonreciprocity by interference of the participating modes coupling phases. We can demonstrate this mathematically by adiabatically eliminating the far-detuned modes, $|\omega - \omega_{c_{0,3}}| \gg g_{uv}, \kappa_{c_{0,3}} $ for $u=1, 2$ \cite{shoreTheoryCoherentAtomic1991, huangLossinducedNonreciprocity2021, yuPredictionAttractiveLevel2019}. By doing this we simplify the original set of equations in Eq. (\ref{eq:sparams}) into a smaller set of equations, involving only equations of motion of the main modes with effective coupling to the far-detuned modes. This yield an effective $\mathbf{S}$ matrix,

\begin{equation}
    \mathbf{S}_{\text{eff}}[\omega] = \mathbf{S}_{\text{bg}} -\mathbf{D}_{\text{eff}}\left[-i\omega \mathbf{I}- \textbf{A}_{\text{eff}}\right]^{-1}\mathbf{B}_{\text{eff}}.
    \label{eq:seff}
\end{equation}

with $\textbf{A}_{\text{eff}}, \mathbf{B}_{\text{eff}}, \mathbf{C}_{\text{bg}}, \mathbf{D}_{\text{eff}}$ defined as,

\begin{equation}
\begin{split}
    \textbf{A}_{\text{eff}} &= \begin{pmatrix}
            -i\tilde\omega_{c_1} & 0 & i\tilde{g}_{10} & i\tilde{g}_{11} \\
            0 & -i\tilde\omega_{c_2} & i\tilde{g}_{20} & i\tilde{g}_{21}\\
           i\tilde{g}_{10}^* & i\tilde{g}_{20}^* & -i\tilde{\omega}_{m_0} - iG_{00} & - iG_{01} \\ 
            i\tilde{g}_{11}^* & i\tilde{g}_{21}^* & - iG_{10} & -i\tilde\omega_{m_1} - iG_{11}
        \end{pmatrix}, \\
\mathbf{B}_{\text{eff}}^\text{T} &= \begin{pmatrix}
            \sqrt{ \gamma_{10} }e^{i \phi_{10}}  & \sqrt{ \gamma_{20} }e^{i \phi_{20}}  & \chi_{0}^{(p=0)} & \chi_{1}^{(p=0)} \\
            \sqrt{\gamma_{11} }e^{i \phi_{11}} & \sqrt{ \gamma_{21} }e^{i \phi_{21}} & \chi_{0}^{(p=1)} & \chi_{1}^{(p=1)} \\
        \end{pmatrix}, \\
        \mathbf{S}_{\text{bg}} &= \begin{pmatrix}
    1 - B_{\text{in,0}}^{\text{out}, 0} & -B_{\text{in,1}}^{\text{out}, 0} \\
    -B_{\text{in,0}}^{\text{out}, 1} & 1 - B_{\text{in,1}}^{\text{out}, 1}
    \end{pmatrix},
     \\
    \mathbf{D}_{\text{eff}} &= \begin{pmatrix}
        \sqrt{\gamma_{10}}e^{-i\phi_{10}} & \sqrt{\gamma_{20}}e^{-i\phi_{20}} & \zeta_{0}^{(p=0)} & \zeta_{1}^{(p=0)} \\
        \sqrt{\gamma_{11}}e^{-i\phi_{11}} & \sqrt{\gamma_{21}}e^{-i\phi_{21}} & \zeta_{0}^{(p=1)} & \zeta_{1}^{(p=1)}
    \end{pmatrix}
        \end{split}
\end{equation}.

Due to the adiabatic elimination of the far-detuned modes we revealed effective coupling terms and indirect crosstalk pathways $\mathbf{S}_{\text{bg}}$ that are both mediated by the far-detuned modes. The effective coupling and crosstalk terms are,

\begin{equation}
\begin{split}
     G_{vv'} &= \sum_{u = 0, 3} \frac{\tilde{g}_{uv}^*\tilde{g}_{uv'}}{\Delta_{c_u}}, \\
    \chi_v^p &= -\sum_{u=0, 3}\frac{\sqrt{\gamma_{up}}e^{i\phi_{up}}\tilde{g}^*_{uv}}{\Delta_{c_u}}, \\
    B_{\text{in, }p_{\text{in}}}^{\text{out, } p_{\text{out}}} &= \sum_{u = 0,3} \frac{i\sqrt{\gamma_{up_{\text{out}}}}e^{-i\phi_{up_{out}}} \sqrt{\gamma_{up_{\text{in}}}}e^{i\phi_{up_{in}}}}{\Delta_{c_u}}, \\
    \zeta_v^p &= \sum_{u=0, 3} \frac{\sqrt{\gamma_{up}}e^{-i\phi_{up}} \tilde{g}_{uv}}{\Delta_{c_u}}, \\
    \label{eq:couplingeff}
\end{split}
\end{equation}

where $G_{vv'}$ are the virtual magnon-magnon coupling terms. $\chi_v^p$ and $\zeta_v^p$ are respectively the coupling of the magnon modes to the probe via the far-detuned modes, and $B_{\text{in, }p_{\text{in}}}^{\text{out, } p_{\text{out}}}$ correspond to the indirect probe-to-probe crosstalk mediated by the far-detuned modes, with $p_{\text{in}}$ and $p_{\text{out}}$ signifying the input and output probe index respectively. 

\begin{figure}
	\centering
	\includegraphics[width=\linewidth]{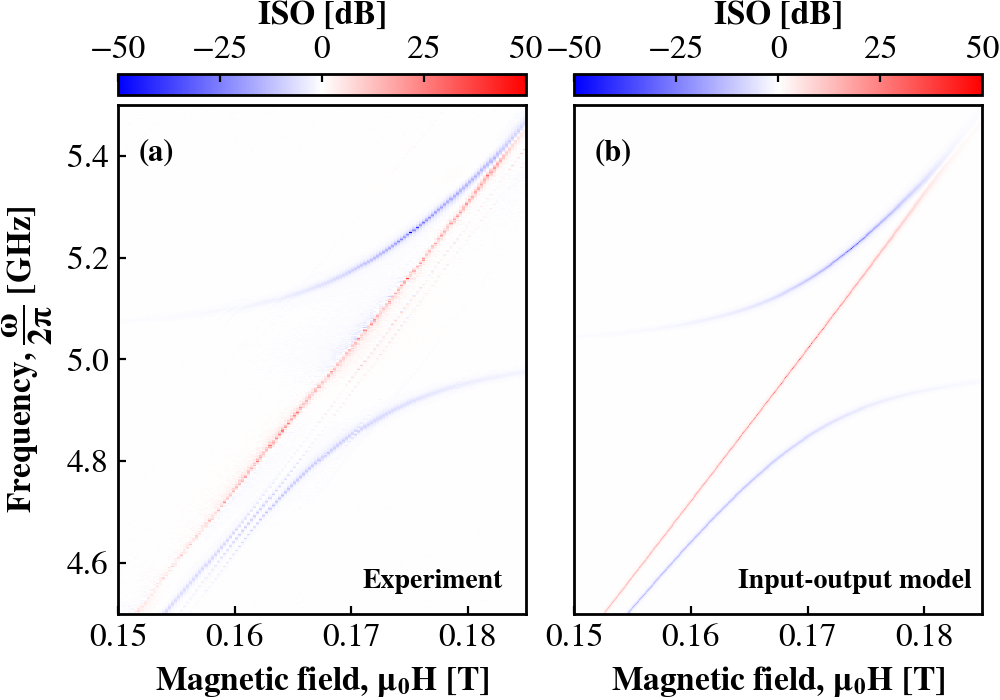}
	\caption{(a) Isolation ratio, frequency and magnetic field resolved ISO map measured in experiment and (b) numerically calculated using the input-output model in Equation 8 and 9. Both ISO map utilize the same color scale in the range of -50 to 50 dB.}
	\label{fig:isocomparison}
\end{figure}

By expanding Eq. (\ref{eq:seff}), we obtain an analytical expression for the isolation ratio (ISO), detailed in Appendix \ref{app:analiso}. Because this expression is algebraically cumbersome, it obscures the specific phase interference mechanisms driving the nonreciprocity. Nevertheless, the fundamental origin of the nonreciprocal transmission can be mathematically traced to the inequality $\mathbf{D}_{\text{eff}} \neq -\mathbf{B}_{\text{eff}}^\dagger$. This nonreciprocity arises directly from the phase-asymmetric effective drive and readout terms, $\chi_v^p$ and $\zeta_v^p$. A comparison between the isolation ratio obtained in experiment and one that is calculated using the input-output model of Eq. (\ref{eq:sparams} and \ref{eq:smatrices}) is provided in Figure \ref{fig:isocomparison}(a) and (b) respectively. We observe good agreement between experiment and the model.

Furthermore, the role of the internal coupling phases (the cavity-magnon coupling phase) in this context is to lift the symmetry by interference with the $\pi$-delayed external coupling phase of $c_0$ and $c_3$. The difference in producing nonreciprocity between DA and PA, lies in the physical phase, $\theta_p$. Since DA has $\theta_p=0$ rad, phase interference with the $\pi$-delayed modes will be the same in both directions, while PA on the other hand has $\theta_p=3.09$ rad which accumulates differently depending on the transmission direction. This shows that to have a complete model that captures the nonreciprocity, it is necessary to include modes that have $\pi$-delayed external coupling phase. To validate the effect of the far-detuned modes and to check for consistency between the effective model (Eq. \ref{eq:seff}) and the full model (Eq. \ref{eq:sparams}), we provide a direct comparison between models with and without the far-detuned modes in Appendix \ref{app:compfar}.

\section{\label{sec:conc}Conclusion}

Bridging, theory, simulation and experiment, we have extended previous studies of coupling phases and further established the importance of these phases in explaining the transmission response of a two port, multi-sphere cavity magnonics system. By explicitly incorporating both internal and external coupling phases within an input–output formalism, we were able to explain the YIG-position-dependent transmission response in terms of the interferences of internal coupling phase. Further, we have shown that the nonreciprocity in the antiresonance frequency as the result of interference effect mediated by far-detuned cavity modes. This work shows that it is crucial to account for coupling phase engineering as a design principle when scaling multimode cavity magnonics system.

\begin{acknowledgments}
We acknowledge that this project has received co-funding from the European Union’s Horizon Europe Research and Innovation Programme under Grant Agreement No. 101126644. This work is also part of the research program supported by the European Union through the European Regional Development Fund (ERDF), as well as by the Ministry of Higher Education and Research and the Brittany region through the CPER SpaceTech DroneTech. Jeremy Bourhill is funded by the Australian Research Council Centre of Excellence for Dark Matter Particle Physics  (CE200100008). We also acknowledge the financial support of the ANR project ICARUS under grant agreement ANR-22-CE24-0008-03. Finally, we thank Bernard Abiven for CNC machining of the cavities used in this study and Kevin Weber for meaningful discussions.
\end{acknowledgments}

\appendix

\section{Ferrimagnetic YIG sphere}

Yttrium iron garnet (YIG) is a ferrimagnetic insulator that possess high spin density, a narrow resonance linewidth with exceptionally low magnon damping and a high Curie temperature \cite{cherepanovSagaYIGSpectra1993, serhaMagneticMaterialsQuantum2025}. These characteristics make YIG suitable for room-temperature experiments. Its properties are well understood and have been extensively used in microwave cavity magnonics experiments \cite{zarerameshtiCavityMagnonics2022, liuTunableCouplingTopology2025, pishehvarOnDemandMagnonResonance2024, zhangMagnonDarkModes2015, bourhillGenerationCirculatingCavity2023, boventerControlCouplingStrength2020}.

Ferromagnets can collectively precess with the same precession frequency due to the large exchange interaction usually present in it. This collective precession allows the macrospin approximation of ferromagnets, where we can consider this quantized excitation of the collective spin as a magnon.

We inserted multiple YIG spheres with diameter $d_{\text{sphere}} = 1$ mm (Ferrisphere, Inc.) into the cavity. The ferromagnetic resonance frequency of the spheres is characterized by $\omega_m / 2\pi = \gamma \mu_0 H$, where $\gamma = 28$ GHz/T is the gyromagnetic ratio and $\mu_0 H$ is the externally applied magnetic field. The saturation magnetization of YIG is 176 T. However in this work, we adjust $\gamma = \gamma_{eff}=29.56$ GHz/T to match experiment.

\section{Tavis-Cummings Coupling model \label{app:Tavis}}

The Tavis-Cummings model treats the interaction of a collective two-level system, assuming rotating wave approximation. The Tavis-Cummings Hamiltonian is expressed as:

\begin{equation}
    \hat{H}_{TC}/\hbar = \tilde{\omega}_c\hat{c}^\dagger \hat{c} + \tilde{\omega}_d \hat{d}^\dagger \hat{d} + g(\hat{c}^\dagger \hat{d} + \hat{c}\hat{d}^\dagger)
\end{equation}

where $\tilde{\omega}_c = \omega_c-i\kappa_c$ is a complex cavity mode angular frequency with damping $\kappa_c$, $\tilde{\omega}_d = \omega_d-i\kappa_d$ is a complex bosonic mode angular frequency for the collective matter excitation with damping $\kappa_d$. The equation of motion bosonic mode operators in the Heisenberg interaction picture is thus,

\begin{align}
    \frac{d}{dt} \hat{\textbf{a}} &= -i\textbf{M}  \cdot \hat{\textbf{a}}, \\ \\
   \textbf{ M} &= \begin{bmatrix}
       \tilde{\omega}_c & g \\
       g & \tilde{\omega}_d
   \end{bmatrix}, \\ \\
    \hat{\textbf{a}}&= \begin{bmatrix}
        \hat{c} \\ \hat{d}
    \end{bmatrix}
\end{align}

By diagonalizing the matrix we can obtain the angular eigenfrequencies,

\begin{equation}
    \omega_{\pm, \text{TC}} = \frac{1}{2} \left[ \tilde{\omega}_c + \tilde{\omega}_d \pm \sqrt{(\tilde{\omega}_c - \tilde{\omega}_d)^2 + 4g^2}\right].
\end{equation}

For more details see \cite{tavisExactSolution$N$MoleculeRadiationField1968, bourcinSpincavitronicsRepulsiveAttractive2024}

\section{\label{app:comsol}Extraction of parameters simulation}

The coupling strengths, coupling phases, azimuthal MW magnetic fields strength, $|H_{\phi}|$ at the probe and the resonance frequency values used for the input-output model are obtained from FE modelling using the RF module in COMSOL Multiphysics\textsuperscript{\tiny\textregistered}. We employ two type of studies, eigenfrequency (EF) and frequency domain (FD) studies. The EF study is used to find the natural resonance frequencies (modes) of a structure. It is a "source-free" simulation, meaning we do not apply an input power or a port excitation. On the other hand, FD study simulates how your device responds to a specific external excitation over a range of frequencies. We must provide an input, such as a Port, Lumped Port, or an Incident Plane Wave. The difference lies in whether we include a drive to the system or not.

The parameter values that are fed into the input-output model are presented in Table \ref{tab:calc_params} with the type of study used to obtain them. Note that we do not exclusively use one type of study to determine the parameter values. This is because extracting one type of value might be more accurate in EF than in FD and vice versa. For example, for parameters where the probe drive is important such as the internal coupling phases and coupling strengths, FD gives a more accurate value giving consistent result when fed into the input-output model. On the contrary, determining the $|H_{\phi}|$ at the probe location is not possible in FD because the probe excitation fields obscure the resonance mode field, giving instead a much more pronounced strength of the field used to excite the cavity. Further, when performing the FD studies, we set the external magnetic field to be far beyond the YIG saturation field (0.176 T). The parameter will be labelled from experiment when we obtain it from the measurement.

\begin{figure}
	\centering
	\includegraphics[width=\linewidth]{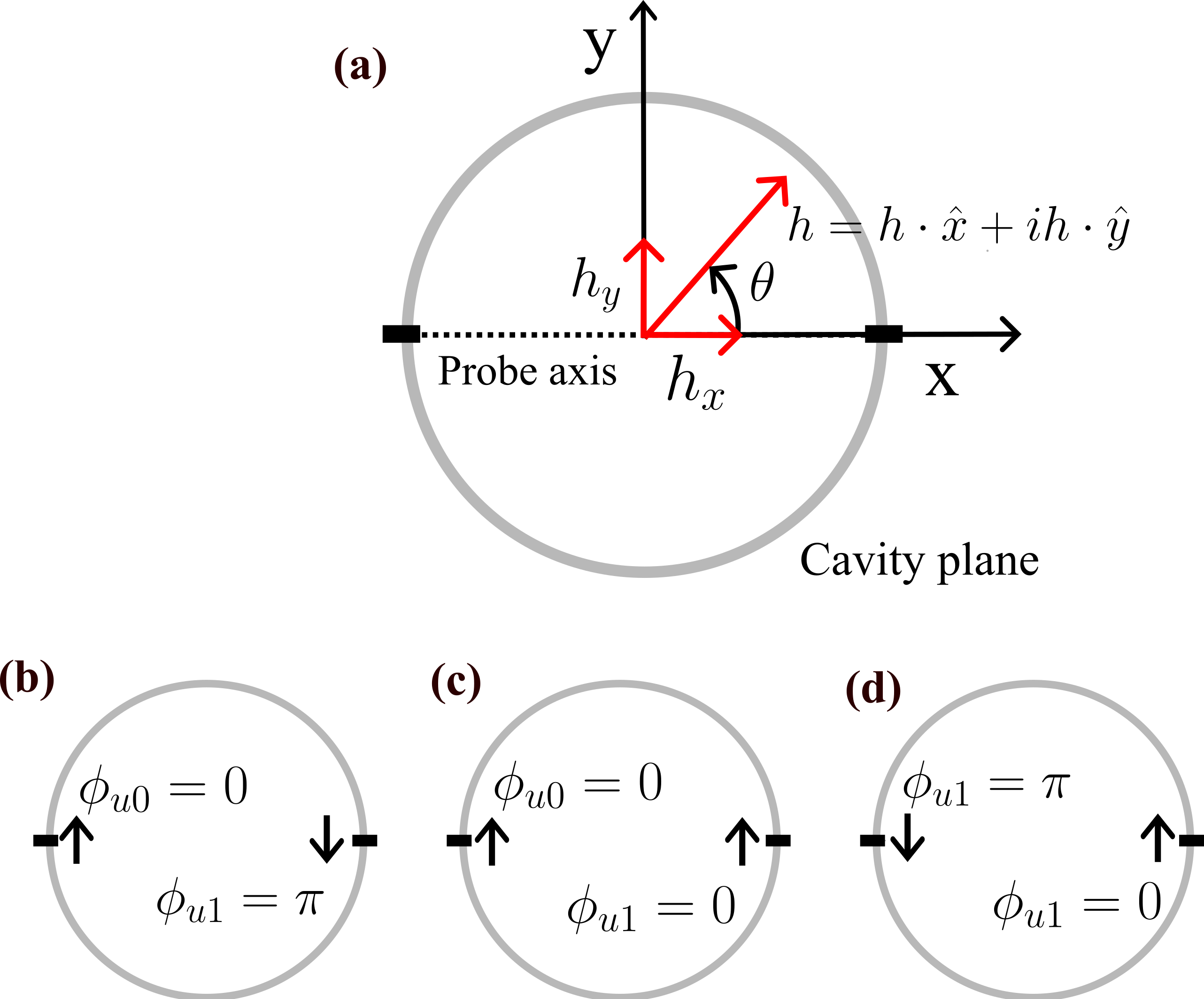}
	\caption{(a) The coordinate system of choice to determine the internal coupling phase, $\theta_{up}$. (b-d) Possible orientation of the MW cavity magnetic field vector and the determined external coupling phase where (b) out-of-phase, (c) in-phase, (d) inversely out-of-phase. Note that inversely in-phase orientation where both $\phi$ are pointing down is equal to (c).}
	\label{fig:arg}
\end{figure}

Since the decay rates of the cavity mode to the probes, $\gamma_{up}$ are proportional to the azimuthal MW magnetic field strength at the probe loop, $|H_\phi|$, we estimate $\gamma_{up}$ as the proportion of a mode's field strength over the sum of all modes field strength,

\begin{equation}
	\gamma_{up}/2\pi = \frac{|H_{\phi, c_u}|}{\sum_{u=0}^3|H_{\phi, c_u}|}\times \text{CF}
\end{equation}

where CF is a coupling factor that has a frequency unit. We adjust CF to reproduce the insertion loss in the experiment, in this work we have $\text{CF} = 10^6$ Hz. Further, the magnon decay rate $\gamma_{m_v}/2\pi$, were determined by taking the 3 dB bandwidth (full-width half-maximum) of the transmission dip corresponding to the FMR, this yield approximately $\gamma_{m_v}=10$ MHz. Note that this value is an effective parameter specific for this experimental setup. It encompasses not only the intrinsic damping parameter of YIG ($\alpha = 4 \times 10^{-4}$) but also additional mechanisms such as inhomogenous broadening and radiative damping present in the system.

We detail the extraction of the coupling strength and coupling phases as follows: 
\begin{enumerate}
	\item The cavity-magnon coupling strengths, $g_{uv}$ were obtained firstly by defining a domain probe that extracts the computed microwave field and calculate the filling factor over the sphere's volume according to Equation 1. Next, we define a global variable probe according to Equation 2 as an expression in COMSOL Multiphysics\textsuperscript{\tiny\textregistered}. The FD or EM study is then run to obtain the coupling strength.
	
	\item The internal coupling phase were calculated using Equation 3, with $x$-axis aligned to the probe axis (see Figure \ref{fig:arg}a). The internal coupling phase is the argument between real and imaginary component of the complex MW cavity magnetic field. The initial reference frame can be chosen arbitrarily since the physically significant property is the relative phase shift between the coupling terms. This can be demonstrated by applying a unitary transformation to the Hamiltonian, see Gardin et al. \cite{gardinManifestationCouplingPhase2023} for more detail.
	
	\item The azimuthal MW magnetic fields strength, $|H_\phi|$ is obtained by defining a point probe at the center point of the probe loop using EF simulation.
	
	\item The external coupling phase were determined from the sign of $H_\phi$ obtained in COMSOL. The sign shows the orientation of the azimuthal magnetic field component at the point probe. Figure \ref{fig:arg}b-d shows the orientation of the magnetic field vector at the point probe and the determined external coupling phase. The external coupling phase transforms through parity inversion between forward and backward transmission direction.
\end{enumerate}

\begin{figure}
	\centering
	\includegraphics[width=\linewidth]{"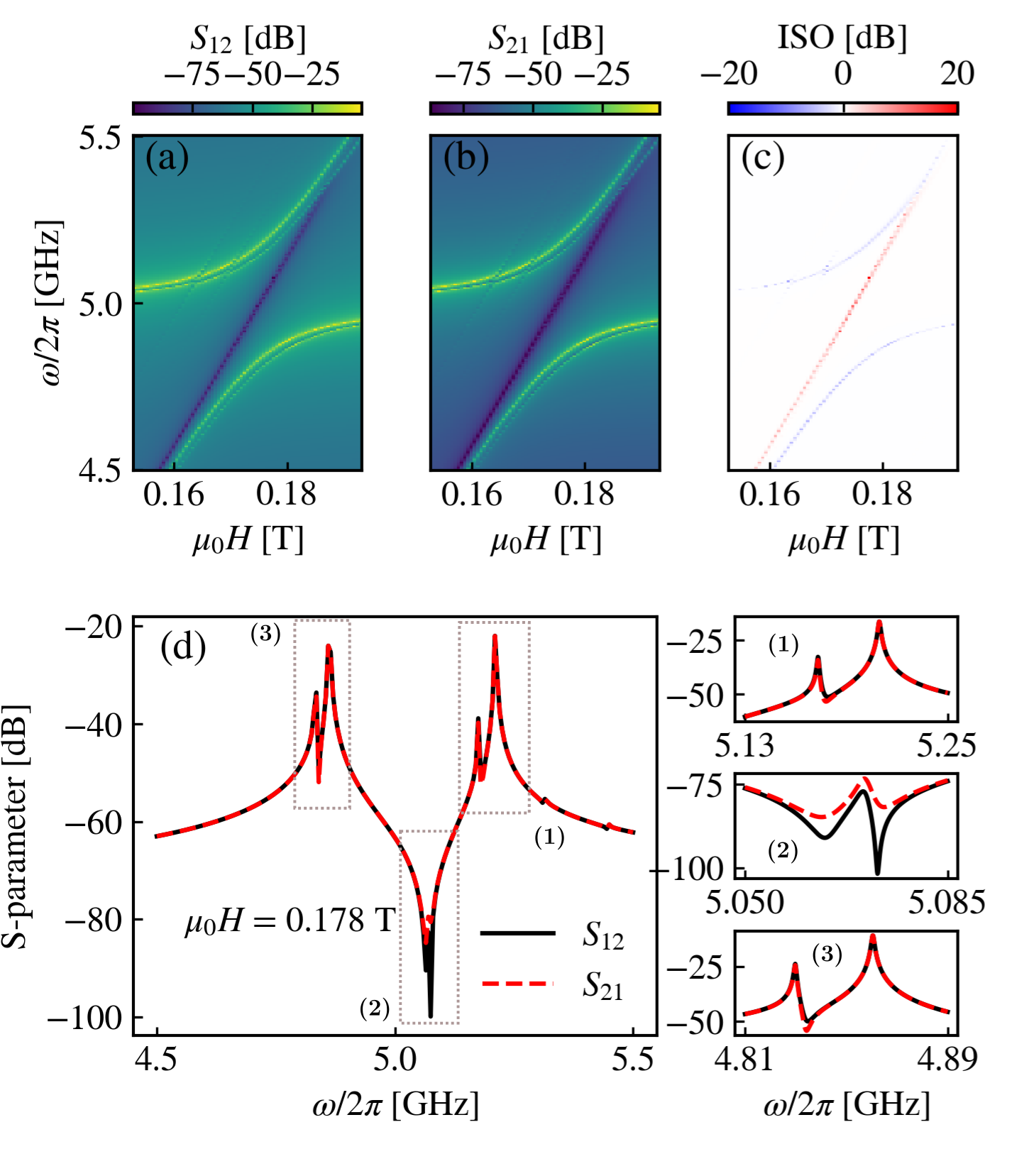"}
	\caption{Simulation results for the parallel arrangement (PA). Color maps of (a) $S_{12}$ [dB], (b) $S_{21}$ [dB], and (c) the isolation ratio (ISO) calculated from simulation data. (d) Frequency line cuts of $S_{12}$ [dB] and $S_{21}$ [dB] evaluated around the saturation field ($\mu_0 H = 0.176\text{ T}$). (d1–d3) Magnified views of the (d1) upper, (d2) ferromagnetic, and (d3) lower resonance features. The magnified views posses enhanced frequency steps, $f_{steps}$ ($f_{steps}=501$). We observe qualitative agreement between simulations, the model and experiment.}
	\label{fig:nonreciprocitysimulation}
\end{figure}

The simulation results agree qualitatively with both the experimental data and the theoretical model, as shown in Figure \ref{fig:nonreciprocitysimulation}(a-c). However, a discrepancy exists in the peak magnitude of the isolation ratio (Figure \ref{fig:nonreciprocitysimulation}(d)), with the numerical model yielding a maximum isolation ratio of $\text{ISO}_{\text{max}} = 19.33\text{ dB}$ at the ferromagnetic resonance frequency. This difference can likely be traced to physical setup imperfections that are absent in the simulation environment.

\section{Transmission response of multisphere cavity magnonics system}

Figure \ref{fig:configs}, \ref{fig:transmission}, and \ref{fig:iso} show the YIG spheres configurations inside the cavity, and its corresponding transmission, and isolation, respectively. We measured up to four YIG sphere that is held in place by a 3D printed sphere holder. In Figure \ref{fig:transmission} different response signature are observed. A result with apparent uncoupled mode can be seen for 1 YIG and diagonal 2 YIG configurations alongside the avoided crossing. On the other hand, the perpendicular 2 YIG configurations, 3 YIG configurations, and 4 YIG configuration showed only avoided crossing. In terms of the isolation, shown in Figure \ref{fig:iso}, 2 YIG perpendicular and 3 YIG configurations show amplified isolation that is located at the antiresonance frequency.

\section{\label{app:unitrans}Unitary transformation}

We follow the approach in \cite{gardinManifestationCouplingPhase2023} to obtain the physical phase, $\theta_p$. By sequentially applying unitary transformations using the unitary matrix $U = e^{i \theta_{uv} a^\dagger a}$ to the Hamiltonian, where $a$ is the bosonic internal mode operator, $a = {c_u, m_v}$.

\begin{equation}
            H \mapsto H' = UHU^\dagger.
\end{equation}

The unitary transformation, transforms $a$ and $a^\dagger$ as,

\begin{align}
    UaU^\dagger &= e^{i \theta_{uv} a^\dagger a} a e^{-i \theta_{uv} a^\dagger a} = ae^{-i\theta_{uv}} \\
    Ua^\dagger U^\dagger &= e^{i \theta_{uv} a^\dagger a} a^\dagger e^{-i \theta_{uv} a^\dagger a}
\end{align}

we obtain a picture where the internal coupling phases are accumulated in one of the coupling pathways. Applying the unitary matrices $U_1 = e^{i\theta_{00}c_0^\dagger c_0}$, $U_2 = e^{i\theta_{10}c_1^\dagger c_1}$, and $U_3 = e^{-i(\theta_{01} - \theta_{00})m_1^\dagger m_1}$ to Eq. (\ref{eq:reduced}) sequentially, yield a Hamiltonian with only one internal coupling phase term $\theta_p = \theta_{21} - \theta_{11} - (\theta_{20} - \theta_{10})$.

\section{\label{app:analiso}Agreement between the full and effective picture}

\begin{figure}[t]
    \centering
    \includegraphics[width=0.8\linewidth]{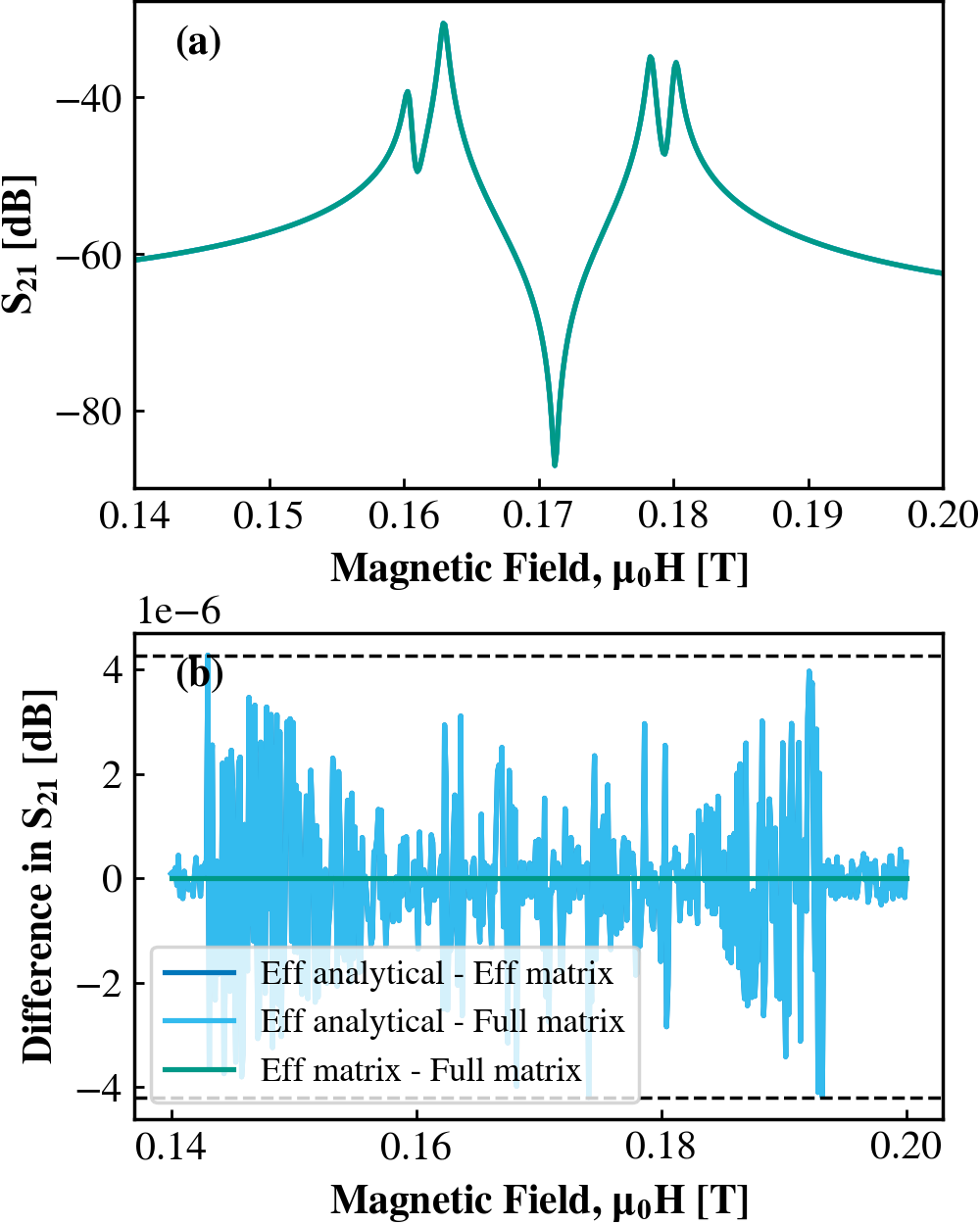}
    \caption{Comparison of the $S_{21}$ transmission obtained from different models. Here the 3 models are overlapping each other. (b) Difference between models. We see that the matrix approach are equivalent, while the analytical approach yield a difference of approximately $\pm 4\times10^{-6}$ dB.}
    \label{fig:agreement}
\end{figure}

In this appendix, we compare the results obtained by solving the full matrix 4 cavity photons/2 magnons description in Eq. (\ref{eq:smatrices}), the effective matrix description of 2 cavity photons/2 magnons, and the analytical $S_{21}$ expression for the effective picture and show that they are equivalent. We can expand Eq. (\ref{eq:couplingeff}) and obtain an analytical $S_{21}$ expression for the effective picture. The expansion yields,

\begin{equation}
S_{21} = \chi_{10} + S_{21,\text{bg}} + R_0 M_0 + R_1 M_1
\label{eq:analyticalS21}
\end{equation}

$M_v$ is the transmission contributions from photons that arrive in Port 2 through $m_0$ and $m_1$ magnon modes. $M_v$ is expressed as,

\begin{align}
M_0 &= i \frac{F_0 D_{11} + F_1 \Sigma_{01}}{D_{00} D_{11} - \Sigma_{01} \Sigma_{10}} \\
M_1 &= i \frac{F_1 D_{00} + F_0 \Sigma_{10}}{D_{00} D_{11} - \Sigma_{01} \Sigma_{10}}
\end{align}

The effective magnon detunings are defined as 

\begin{align}
    D_{00} &= \Delta_{m_0} - \Sigma_{00} \\
    D_{11} &= \Delta_{m_1} - \Sigma_{11},
\end{align}

with virtual magnon self-energy and magnon-magnon coupling terms,

\begin{align}
\Sigma_{vv} &= \sum_{u} \frac{g_{u v}^2}{\Delta_{c_u}} \quad \text{for } v \in \{0, 1\} \\
\Sigma_{vv'} &= \sum_{u} \frac{g_{u v} g_{u v'} e^{-i(\theta_{cv} - \theta_{cv'})}}{\Delta_{c_u}} \quad \text{for } v \neq v'
\end{align}

Further, the effective drive, $F$ and readout, $R$ terms are

\begin{align}
F_v &= - \sum_{u} \frac{g_{uv} \sqrt{\gamma_{u0}}}{\Delta_{c_u}} e^{-i\theta_{uv} + i\phi_{u0}} \\
R_v &= \sum_u \frac{g_{uv} \sqrt{\gamma_{u1}}}{\Delta_{c_u}} e^{i\theta_{uv} - i\phi_{u1}}
\end{align}

while the virtual crosstalk terms are,
\begin{equation}
S_{21,\text{bg}} = -i \sum_{u} \frac{\sqrt{\gamma_{u0} \gamma_{u1}}}{\Delta_{c_u}} e^{i(\phi_{u0} - \phi_{u1})}
\end{equation}

The complex bare detunings, $\Delta_{c_u}$ and $\Delta_{m_v}$ in the grouped terms are defined as,
\begin{align}
\Delta_{c_u} &= \omega - \omega_{c_u} + i \frac{\kappa_{c_u}}{2} \\
\Delta_{m_v} &= \omega - \omega_{m_v} + i \frac{\kappa_{m_v}}{2}
\end{align}
where $\omega = 2\pi f$ is the driving frequency. For $u \in \{0, 1, 2, 3\}$. In Figure \ref{fig:agreement}, we plot numerically S21 transmission at $\mu_0 H = 0.170$ T result using the full 4 cavity modes/2 magnons system using the matrix Eq. (\ref{eq:smatrices}), the effective 2 cavity modes/2 magnon modes matrix in Eq. (\ref{eq:couplingeff}), and the analytical expression Eq. (\ref{eq:analyticalS21}). The results in Figure \ref{fig:agreement}(a) are overlapping with each other. We plot the difference in $S_{21}$ in Figure \ref{fig:agreement}(b). The full and effective matrices are equivalent, shown by the flat difference, while the difference between the analytical and matrix approach yield approximately $\pm 4\times10^{-6}$ dB difference which is very small compared to the scale of transmission.

\section{Inclusion of the far-detuned modes \label{app:compfar}}

\begin{figure}
    \centering
    \includegraphics[width=\linewidth]{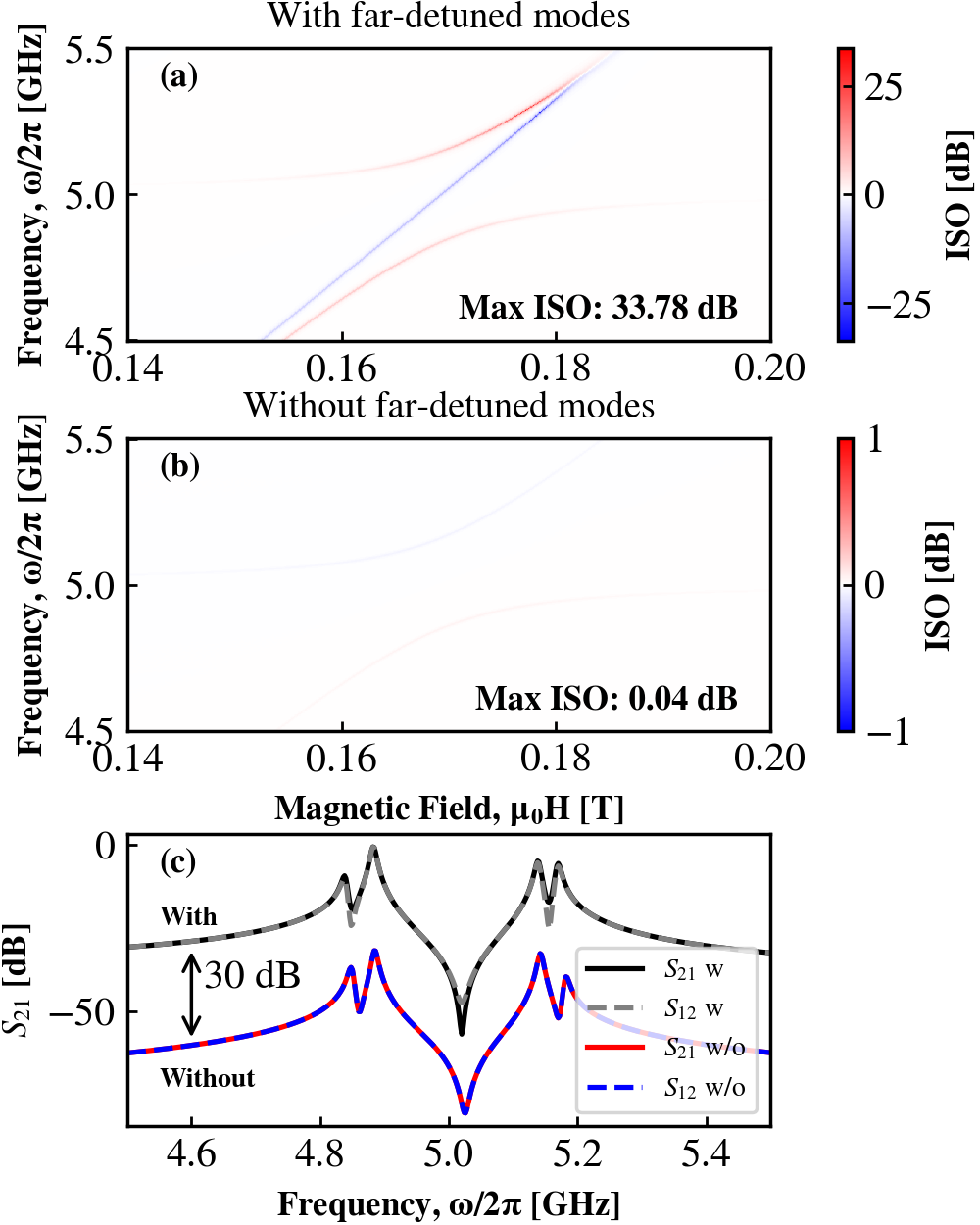}
    \caption{ISO map of models (a) with the inclusion of far-detuned modes, and (b) without the far-detuned modes. (c) A frequency line cut of both models at $\mu_0 H = 170$ mT. The model with external coupling phase included is shifted by +30 dB}
    \label{fig:farmodescomp}
\end{figure}

We argue in the main text that the far-detuned modes are responsible for the nonreciprocal transmission. In Figure \ref{fig:farmodescomp}(a) and (b), respectively, we compare the model with the far-detuned modes included and removed. The removal is done by removing the $c_0$ and $c_3$ modes, setting the coupling strengths $g_{0v} = g_{3v} = 0$, the external dissipation $\gamma_{0p} = \gamma_{3p} = 0$ and $\tilde{\omega}_{c_0} = \tilde{\omega}_{c_3} = 0$.

By including the far-detuned modes into the model we were able to reproduce the nonreciprocity. In Figure \ref{fig:farmodescomp}, we can observe the nonreciprocity at the antiresonances.

\begin{table*}
	\caption{Relevant parameters extracted from COMSOL Multiphysics\textsuperscript{\tiny\textregistered} eigenfrequency (EF) and frequency domain (FD) studies. In the main text, we used parameters from frequency domain studies.}
	\centering
	\begin{ruledtabular}
		\begin{tabular}{cccccccccccc}
			Config. & Mode index, $u$ & $\omega_{c_{u}}/2\pi$ [GHz] & $g_{u0}/2\pi$ [MHz] & $g_{u1}/2\pi$ [MHz] & $\gamma_{c_u}/2 \pi$ [MHz] & $\gamma_{m_v}/2\pi$ [MHz] & $\frac{\gamma_{u0}}{2\pi}, \frac{\gamma_{u1}}{2\pi}$ [MHz] \\
			\hline
			DA  & $0$ & $2.6586$    & $4.3587$  & $4.3716$ & $5.10$ & $10$ & $0.4981$   \\
			& $1$  & $5.0007$   & $90.445$  & $90.445$ & $9.28$ & $10$ & $0.3386$  \\
			& $2$ & $5.0065$    & $90.503$  & $90.503$ & $9.28$ & $10$ & $0.0023$  \\
			& $3$  & $6.2957$   & $114.67$  & $114.62$ & $9.30$ & $10$ & $0.1610$  \\
			\hline
			PA  & $0$ & $2.6586$    & $3.4615$  & $3.4624$ & $5.10$ & $10$ & $0.4981$  \\
			& $1$ & $5.00071$   & $78.526$  & $113.91$ & $9.28$ & $10$ & $0.33865$ \\
			& $2$ & $5.0065$    & $112.12$  & $110.63$ & $9.28$ & $10$ & $0.0023$   \\
			& $3$ & $6.2958$    & $132.91$  & $133.01$ & $9.30$ & $10$ & $0.1610$  \\
			\hline
			Study & & EF & FD & FD & Measured & Measured & EF 
		\end{tabular}
	\end{ruledtabular}
	\begin{ruledtabular}
		\begin{tabular}{cccccccccccc}
			Config. & Mode index, $u$ & $\theta_{u0}$ [rad] & $\theta_{u1}$  [rad] & $\phi_{u0}$ [rad] &$\phi_{u1}$ [rad] & $\gamma$ [GHz/T] &  &    \\
			\hline
			DA  & $0$ & $0.79292$   & $-2.3442$  &  $0$    & $\pi$ & $29.56$ & &  \\
			& $1$ & $-0.73956$  & $-0.73948$ &  $0$    & $0$   & $29.56$ & &   \\
			& $2$ & $2.2635$    & $2.2635$   &  $0$    & $0$   & $29.56$ & &   \\
			& $3$ & $-0.78537$  & $2.3562$   &  $\pi$  & $0$   & $29.56$ & &   \\
			\hline
			PA  & $0$ & $2.3624 $  & $-2.3238$  & $0$     & $\pi$ & $29.56$ & &   \\
			& $1$ & $0.72727$   & $-0.7164$   & $0$     & $0$   & $29.56$ & &   \\
			& $2$ & $0.84963$   & $2.2938$   & $0$     & $0$   & $29.56$ & &   \\
			& $3$ & $-0.785072237$   & $0.785453493$   & $\pi$   & $0$   & $29.56$ & &  \\
			\hline
			Study & & FD & FD & EF & EF & Measured & & & 
		\end{tabular}
		\label{tab:calc_params}
	\end{ruledtabular}
\end{table*}

\begin{figure*}
    \centering
    \includegraphics[width=\linewidth]{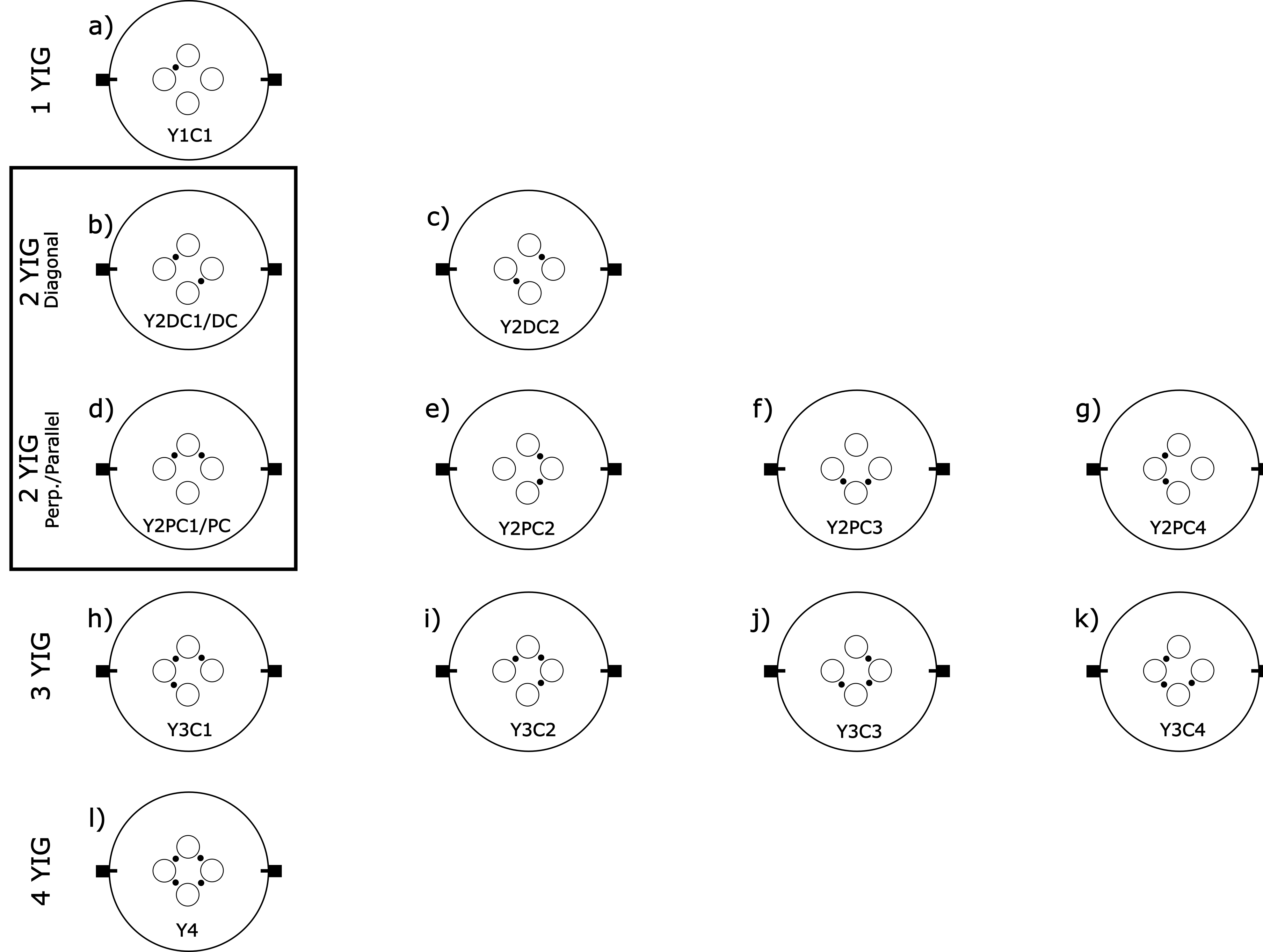}
    \caption{Schematics of the YIG position inside the cavity resonator, the configuration labels match the transmission and isolation response in Figure \ref{fig:transmission} and \ref{fig:iso}. The configurations inside the box are discussed in the main text.}
    \label{fig:configs}
\end{figure*}

\begin{figure*}
    \centering
    \includegraphics[width=\linewidth]{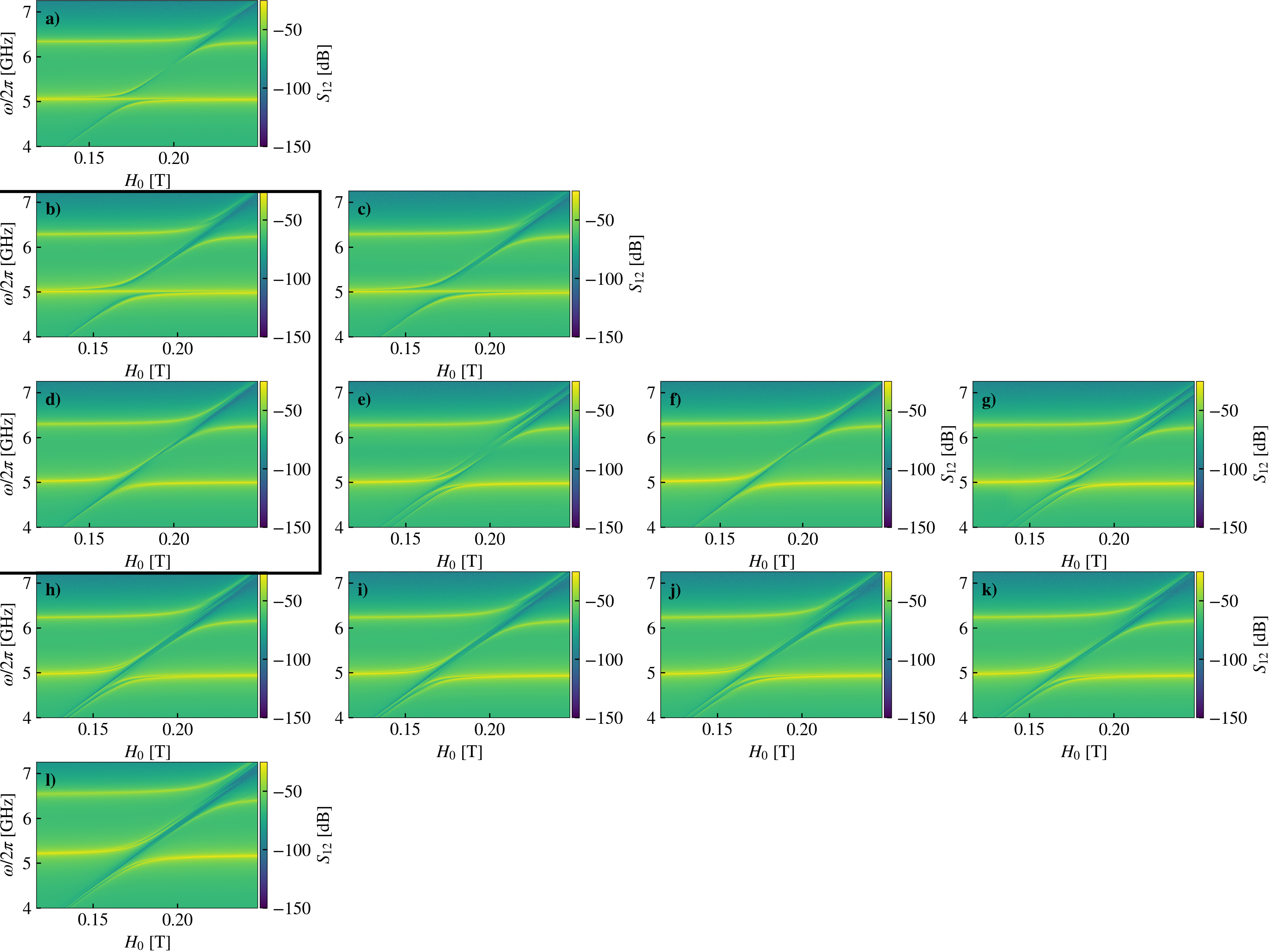}
    \caption{Transmission response of the cavity magnonics system for configurations shown in Figure \ref{fig:configs}. The responses inside the box are discussed in the main text}
    \label{fig:transmission}
\end{figure*}

\begin{figure*}
    \centering
    \includegraphics[width=\linewidth]{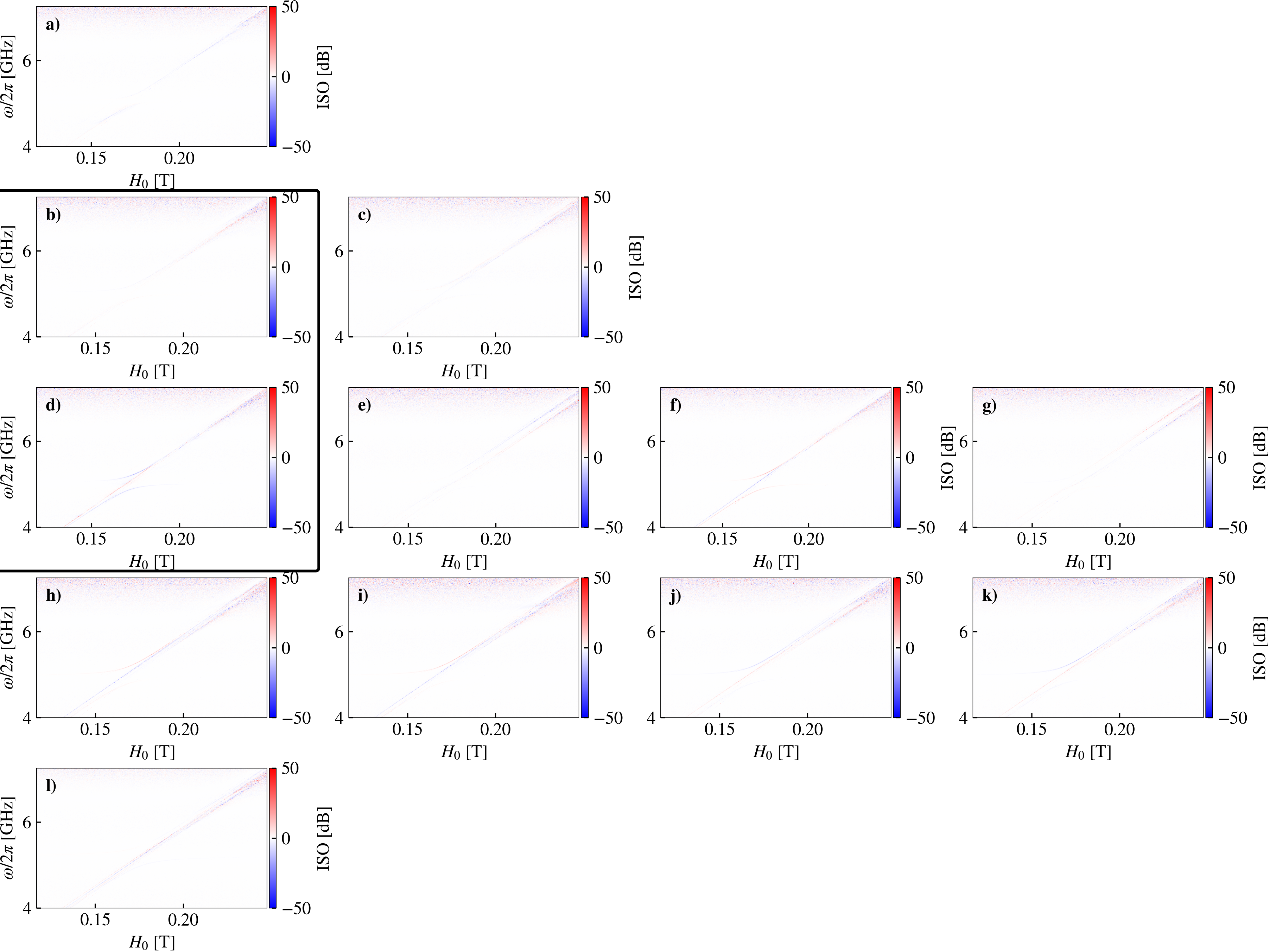}
    \caption{Isolation response of the cavity magnonics system for configurations shown in Figure \ref{fig:configs}. The responses inside the box are discussed in the main text}
    \label{fig:iso}
\end{figure*}

\clearpage

\bibliography{references}

\end{document}